\documentclass[journal]{IEEEtran}
\usepackage{amsmath,amsfonts}
\usepackage{algorithmic}
\usepackage{algorithm}
\usepackage{array}
\usepackage[caption=false,font=normalsize,labelfont=sf,textfont=sf]{subfig}
\usepackage{textcomp}
\usepackage{stfloats}
\usepackage{url}
\usepackage{verbatim}
\usepackage{graphicx}
\usepackage{cite}
\usepackage{orcidlink}
\usepackage[dvipsnames]{xcolor}
\usepackage{comment}
\usepackage{hyperref}
\usepackage{tikz}
\usepackage[most]{tcolorbox}
\usepackage{rotating}
\usepackage{cleveref}[2012/02/15]%
\usepackage{tabularray}
\usepackage{enumitem}
\usepackage[normalem]{ulem}

\UseTblrLibrary{booktabs}
\UseTblrLibrary{varwidth}

\definecolor{newtext}{rgb}{0.0, 0.0, 0.7}
\definecolor{removed}{rgb}{0.7, 0.0, 0.0}

\newcommand{\added}[1]{#1}
\newcommand{\removed}[1]{}

\begin{document}

\title{On the Illusion of Success: An Empirical Study of Job Reruns and Silent Failures in Industrial CI}

\author{
Henri A\"idasso \orcidlink{0009-0004-1625-0159},~\IEEEmembership{Student Member,~IEEE,}
Francis Bordeleau \orcidlink{0000-0001-7727-3902},~\IEEEmembership{Member,~IEEE,}
\break and
Ali Tizghadam \orcidlink{0000-0002-0898-3094},~\IEEEmembership{Member,~IEEE}

}

\maketitle

\begin{abstract}
 Reliability of pipeline outcomes is a cornerstone of effective Continuous Integration (CI) and Continuous Deployment (CD) of software. Yet in practice, developers often struggle with non-deterministic issues in the code or CI/CD environment, which undermine trust in CI/CD pipeline results. When faced with such unexpected outcomes, developers tend to repeatedly rerun jobs hoping for true success, but this practice is known to increase CI costs and reduce productivity. While recent studies have focused on intermittent job failures, no prior work has investigated \removed{silent failures, where pipeline jobs are marked as successful but fail to complete all or part of their tasks.}\added{CI jobs that are marked as successful but fail to complete their intended tasks (i.e., silent failures)}. Such issues often go unnoticed, creating an illusion of success with detrimental consequences such as bugs escaping into production, and prompting the rerun of successful jobs. This paper presents the first empirical study of silent failures through the practice of rerunning successful jobs. An analysis of 142,387 jobs across 81 industrial projects shows that 11\% of successful jobs are rerun, with 35\% of these reruns occurring after more than 24 hours. Using mixed-effects models on 32 independent variables (AUC of 85\%), we identified key factors associated with reruns of successful jobs, notably testing and static analysis tasks, scripting languages like Shell, and developers’ prior rerun tendencies. A further analysis of 92 public issues revealed 11 categories of silent failures aligning with these factors, the most frequent being artifact operation errors, caching errors, and ignored exit codes. Overall, our findings provide valuable insights into the circumstances and causes of silent failures to raise awareness among teams, and present solutions to improve CI reliability.

\end{abstract}


\newcommand{\rqone}{\textbf{RQ\textsubscript{1}.} How often do developers rerun successful jobs? What is the impact of rerunning successful jobs?}
\newcommand{\rqoneone}{\textbf{RQ1.1.} How often do developers rerun jobs?}
\newcommand{\rqonetwo}{\textbf{RQ1.2.} What is the impact and cost of rerunning successful jobs?}
\newcommand{\rqtwo}{\textbf{RQ\textsubscript{2}.} What are the most important circumstances under which developers rerun successful jobs?}
\newcommand{\rqthree}{\textbf{RQ\textsubscript{3}.} What types of failures occur silently within jobs marked as successful?}
\newlength\MAX  \setlength\MAX{6mm}
\newcommand*\Chart[1]{\strut{
    \rlap{\textcolor{black!20}{\rule{\MAX}{2ex}}}\rule{#1\MAX}{2ex}
    }
}

\definecolor{greenstatus}{RGB}{0, 204, 150}

\newcommand{\DrawPercentageBar}[1]{%
  \begin{tikzpicture}
    \fill[color=black!50]   (0.0 , 0.0) rectangle (#1*6ex , 1.5ex );
    \fill[color=black!15] (#1*6ex  , 0.0) rectangle (6.0ex, 1.5ex);
  \end{tikzpicture}%
}

\newcommand{\DrawPercentageBarRed}[1]{%
  \begin{tikzpicture}
    \fill[color=red!60]   (0.0 , 0.0) rectangle (#1*6ex , 1.5ex );
    \fill[color=black!15] (#1*6ex  , 0.0) rectangle (6.0ex, 1.5ex);
  \end{tikzpicture}%
}

\newcommand{\DrawPercentageBarGreen}[1]{%
  \begin{tikzpicture}
    \fill[color=greenstatus]   (0.0 , 0.0) rectangle (#1*6ex , 1.5ex );
    \fill[color=black!15] (#1*6ex  , 0.0) rectangle (6.0ex, 1.5ex);
  \end{tikzpicture}%
}

\newtcolorbox{fancyquote}[1][]{%
  enhanced,
  sharp corners,
  colback=gray!5,
  colframe=gray!70,
  coltitle=black,
  fonttitle=\bfseries,
  left=6pt,
  right=6pt,
  top=6pt,
  bottom=6pt,
  boxrule=0pt,
  frame hidden,
  borderline west={2pt}{0pt}{orange!80}, %
  #1
}

\definecolor{lightgray}{gray}{0.97}

\lstset{
    basicstyle=\ttfamily\scriptsize\color{black!90},
    frame=lines, %
    breaklines=true,
    columns=fullflexible,
    escapeinside={<@}{@>},
    frameround=ffff,
    framerule=.5pt,
    numbers=left,
    numberstyle=\ttfamily\scriptsize,
    numbersep=8pt,
    xleftmargin=15pt,                   %
    xrightmargin=0pt,                  %
    framexleftmargin=15pt,              %
    framexrightmargin=0pt,             %
    columns=fullflexible,
    linewidth=\linewidth, 
    showstringspaces=false
}

\tcbset{
  rqbox/.style={
    enhanced,
    attach boxed title to top left={xshift=1em,yshift=-\tcboxedtitleheight/2},
    colback=black!5!white,
    colframe=black!70!white,
    colbacktitle=red!80!black,
    drop shadow={black!50!white},
    coltitle=white,
    arc=4pt,
    outer arc=4pt,
    left=6pt,
    right=6pt,
    top=10pt,
    bottom=4pt,
    title=#1,
    boxed title style={
        size=small,
        colback=gray,
        boxrule=1.5pt,
        top=.7pt,
        bottom=.2pt,
        left=8pt,
        right=8pt,
    }
  }
}

\crefformat{footnote}{#2\footnotemark[#1]#3}

\section{Introduction}
\label{sec:introduction}

\IEEEPARstart{M}{odern} software organizations widely use Continuous Integration and Continuous Deployment (CI/CD) pipelines to automate the packaging, testing, and deployment of their software products \cite{hilton_usage_2016}. Also referred to as build \cite{aidasso_build_2025}, the CI (and CD) pipeline is automatically triggered when developers submit their code changes to the central code repository \cite{jin_what_2021}. The \removed{build}\added{pipeline} typically involves the execution of several jobs, including compilation, static code analysis, unit and integration testing, binary or container image creation, and deployment of the new software version into production.
 On the one hand, developers are encouraged to trigger the \removed{build}\added{pipeline} multiple times per day \cite{shahin_continuous_2017}, enabling organizations to swiftly respond to market changes. On the other hand, \removed{build}\added{pipeline} failures provide developers with fast feedback on code integration errors (e.g., bugs, test failures, code smells, and merge conflicts), enabling them to detect and fix these issues early in the software development process \cite{hilton_usage_2016, elazhary_uncovering_2022}. As a result, CI helps organizations deliver software faster and more often to their end-users while maintaining high-quality standards.

Reliable \removed{build}\added{pipeline} outcomes are fundamental to achieving the benefits of CI/CD. 
For our industrial partner TELUS, a leading telecommunications company, automated \removed{build}\added{pipeline}s ensure frequent and rapid release of critical network softwarization products. These products include software-defined networks (SDNs) \cite{noauthor_software-defined_nodate}, which offer a new networking paradigm where software-based controllers are used to dynamically manage large network traffic. In fact, SDNs are used in a wide range of fields where high and stable connectivity plays a central role, such as telehealth, internet access, smart cities, connected agriculture, and television \cite{telus_telus_2021}. In such a context, ensuring that \removed{build}\added{pipeline}s provide developers with reliable outcomes is of the utmost importance to prevent bugs from escaping into production while maintaining productive teams and cost-effective CI/CD processes.

In practice, unfortunately, \removed{build}\added{pipeline} outcomes do not always accurately reflect whether the specified tasks were truly completed or had failed. In fact, prior studies \cite{lampel_when_2021, olewicki_towards_2022, aidasso_diagnosis_2025, ghaleb_studying_2019} indicated a growing number of \removed{build}\added{pipeline} failures that do not indicate any issue in the submitted code changes, as developers would expect. 
Instead, intermittent (i.e., irregular and non-deterministic) job failures arise from instabilities in the CI environment and various other anomalies such as non-deterministic test cases, networking issues, concurrency problems, and overloaded servers \cite{ghaleb_studying_2019, aidasso_diagnosis_2025}. This problem of intermittent job failures (including not only flaky tests but also CI environment issues) has been more frequently reported in industrial settings \cite{lampel_when_2021, olewicki_towards_2022, aidasso_diagnosis_2025}, where they hinder developers' interpretation of \removed{build}\added{pipeline} failures, disrupting the software development lifecycle with added costs and delays.

Compounding the problem of intermittent job failures is the phenomenon of \textit{silent failures}, where a \removed{build}\added{pipeline} job reports success (exit code 0, green check mark) despite failing to complete its intended tasks, fully or partially. These silent job failures typically remain undetected until a developer notices a related problem at a later stage, such as a missing job artifact or an expected software version absent from production. One such example is documented in a reported issue within the open-source GitLab Runner project\footnote{\url{https://gitlab.com/gitlab-org/gitlab-runner/-/issues/2291}}, where the cache upload step failed silently in a job appearing successful, without any trace of error or warning in the logs. 

As a consequence, silent failures cause delayed feedback to developers, erode their trust in job successes, and lead to ineffective CI processes. Indeed, when faced with unclear or unexpected \removed{build}\added{pipeline} outcomes, developers often rerun the suspected jobs multiple times (see Fig.~\ref{fig:restart_suite}) \textit{just to be sure} \cite{olewicki_towards_2022} and in the hope that the job will ultimately run as expected \cite{olewicki_towards_2022, aidasso_diagnosis_2025}. 
Such frequent reruns place additional strain on already limited CI server resources, increase developer waiting times, and quickly escalate CI costs \cite{aidasso_diagnosis_2025, olewicki_towards_2022, aidasso_build_2025}. 
Prior work \cite{durieux_empirical_2020, olewicki_towards_2022, aidasso_diagnosis_2025, aidasso_efficient_2025} studied the practice of rerunning jobs (both in open-source and industrial contexts), focusing on intermittent job failures, where explicitly failed jobs are rerun until they pass. \removed{In contrast, silent failures in seemingly successful jobs have been overlooked in the existing literature, most certainly due to the challenge of identifying and investigating them without explicit error traces.}\added{More recently, Maipradit et al.~\cite{maipradit_repeated_2024} and Brus et al.~\cite{brus_rechecking_2025} studied rechecks of \emph{failing} CI runs in OpenStack, modeling the likelihood of a recheck resolving a failure. In contrast, the complementary phenomenon of silent failures, where jobs \emph{appear successful} but fail to complete their intended tasks, has received little attention, most likely due to the challenge of identifying them without explicit error traces.}

Therefore, silent failures constitute a critical, underaddressed threat to the reliability of CI processes. Indeed, they (1) undermine automation goals by increasing manual interventions and hindering reproducibility, (2) cause delayed detection of bugs and release issues, and (3) lead to wasteful reruns of \removed{build}\added{pipeline} jobs. Hence, there is a strong need for our industrial partner to understand the circumstances under which silent failures occur, in order to raise awareness among teams and support their effective identification and resolution. These findings will also benefit practitioners aiming to improve the reliability of their CI processes and open new research directions on CI reliability for researchers.

 To this end, we first assess the prevalence and cost of rerunning (successful) jobs by analyzing 142,387 jobs collected across 81 projects at TELUS and spanning 6 years. Then, we investigate the most important factors associated with the rerun of successful jobs to better understand the circumstances under which such reruns occur. Finally, we examine 92 publicly available issues related to silent failures to identify the most common causes and existing solutions. Specifically, we answer the following research questions (RQs):

\textbf{\rqone} In the studied projects, about 11\% of successful jobs (11\%) have been rerun. These reruns account for more than half (51\%) of all reruns without code changes, even more than reruns of failed jobs (44\%). While reruns of failed jobs typically occur within 15 minutes, over 35\% of the reruns of successful jobs are triggered after 24 hours, with delays even exceeding 72 hours. Moreover, reruns of successful jobs consume nearly half (48\%) of the total server time spent on all job reruns. These findings highlight the prevalence of silent issues prompting reruns and their substantial impact on CI costs.

\textbf{\rqtwo} We studied the influence of 32 independent variables, grouped into three families, on the likelihood of rerunning a successful job using mixed-effects logistic regression models. The results revealed influential factors associated with a higher probability across all three families: job execution context (e.g., code or image build, testing, and static analysis jobs), project and its history (e.g., projects dominated by scripting languages like Shell, or with a larger number of commits), and the job author's recent tendency to rerun jobs. Inversely, some factors are significantly associated with a lower probability of rerunning successful jobs, such as executions on feature branches where developers work in isolation, projects with higher \removed{build}\added{pipeline} failure ratios, and authorship from more experienced developers.

\textbf{\rqthree} We manually examined 92 public issues where practitioners reported cases of silent failures in jobs appearing successful. We identified 11 themes, with the top three most common being artifact operation errors (28\%), caching errors (23\%), and ignored non-zero exit codes (18\%). \removed{These themes align with the identified influential factors. In fact, testing and static analysis jobs often handle artifacts (e.g., test reports) and caches (e.g., dependencies), while ignored exit codes are common in scripting-heavy jobs (e.g., Shell)}\added{A cross-tabulation of themes against job types (coded from issue descriptions) shows that build and testing jobs concentrate in these three most frequent themes, consistent with their higher rerun odds in RQ2. Security scanning failures map exclusively to security scan jobs, and post-deployment issues exclusively to deployment jobs, further confirming the connection between the quantitative and qualitative findings}. Finally, we summarized the practical solutions proposed for these issues to facilitate their reuse by practitioners.

\textbf{Paper Organization.} The remainder of this paper is organized as follows. Section~\ref{sec:background} presents the background and motivation of this study. Section~\ref{sec:study_design} describes the study design. Section~\ref{sec:findings} presents the findings, while Section~\ref{sec:implications} discusses their implications. Furthermore, Section~\ref{sec:related} reviews related work, Section~\ref{sec:threats} outlines potential threats to validity, and Section~\ref{sec:conclusion} summarizes and concludes this paper.

\textbf{Replication Package.} We make publicly available a {replication package} \cite{aidasso_replication_2025} that includes: \textbf{(1) Datasets} of 142,387 jobs from 81 industrial projects, 198,146 jobs from 8 popular open-source projects, and 92 identified public issues related to silent failures; \textbf{(2) Source Code},  notebooks, along with intermediate results to enable verification and replication of our study; and an \textbf{(3) Online Appendix} providing additional study details and results of our replication study on the 8 open-source projects, supporting the external validity of our findings.

\section{Background and Motivation}
\label{sec:background}

\begin{figure}
  \begin{center}
      \includegraphics[width=0.65\linewidth]{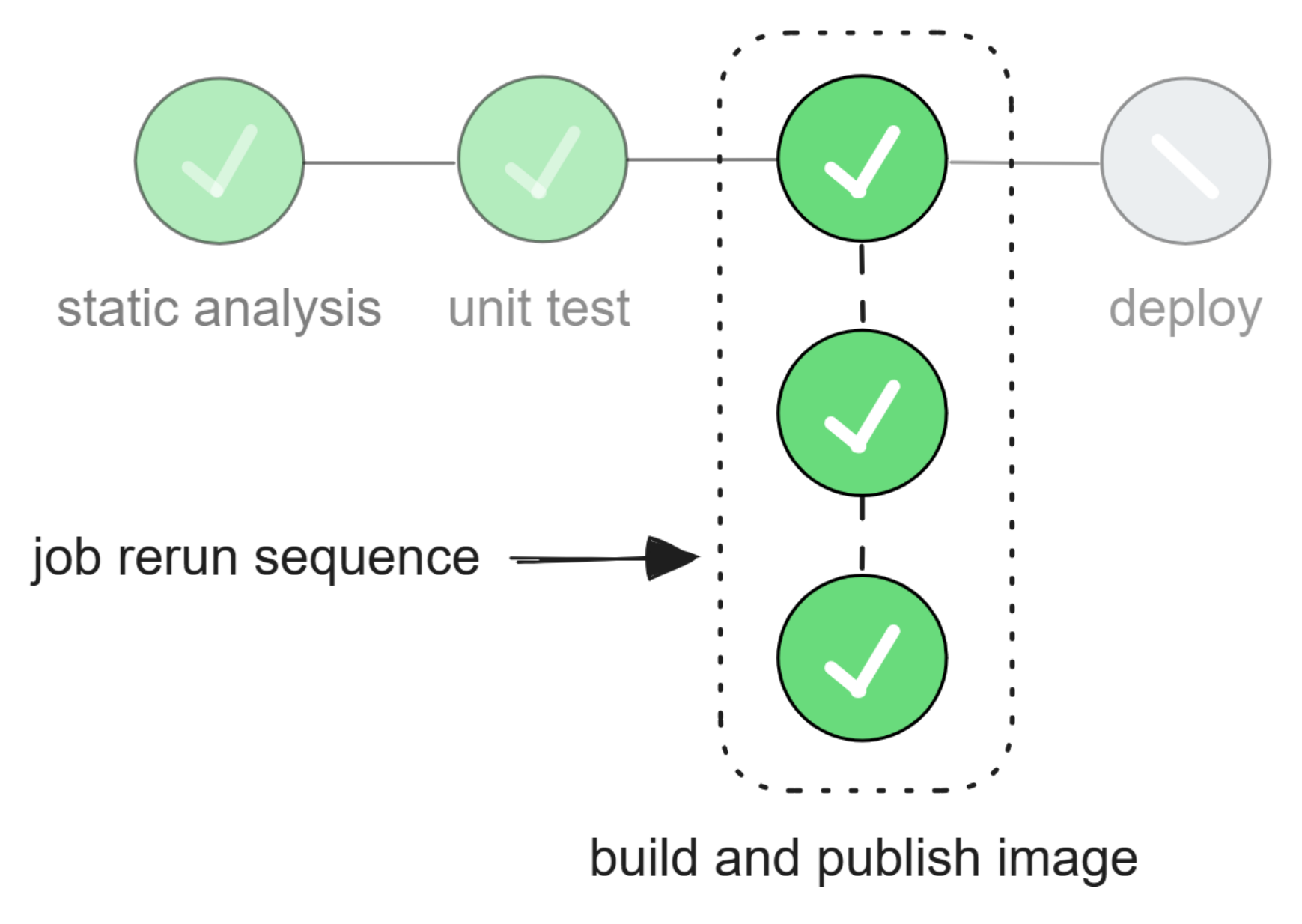}
  \end{center}
\caption{Illustration of a typical CI pipeline of four jobs, highlighting the rerun sequence of the third job responsible for building and publishing the container image. This job is re-executed twice due to silent failures, resulting in three total executions of this same job without any code or \removed{build script}\added{CI configuration file} change.}
\label{fig:restart_suite}
\end{figure}

\textbf{Continuous Integration.} Continuous Integration (CI) and Continuous Deployment (CD) are complementary software engineering practices that consist of automating all the steps that code changes undergo to be safely delivered to end-users \cite{hilton_usage_2016, aidasso_build_2025}. In practice, developers are encouraged to trigger a pipeline of jobs, referred to as the CI/CD pipeline (or build) \cite{aidasso_build_2025}, by submitting small code changes to the version control system. On code changes submission, the pipeline execution is handled by dedicated tools called CI systems, such as GitLab CI and GitHub Actions. These platforms provide convenient SaaS services that make it easy to \removed{automate builds and deployments}\added{configure and execute CI/CD pipelines}. However, an ever-growing number of organizations choose to host their own versions of CI systems inside Virtual Private Networks (VPNs) to ensure the security and protection of proprietary software assets. 

\added{Throughout this paper, we use the following CI terminology consistently. A \textbf{pipeline} (also referred to as a \textbf{build} in the literature~\cite{aidasso_build_2025}) is a set of jobs triggered by a code change (e.g., a commit or merge request). A \textbf{job} is a single execution unit within a pipeline that performs a specific task (e.g., testing, linting, deployment). A \textbf{CI configuration file} (e.g., \texttt{.gitlab-ci.yml}), sometimes called a \textbf{build script} in the literature~\cite{aidasso_build_2025}, defines the pipeline structure, jobs, and their execution rules. Finally, a \textbf{rerun} refers to the re-execution of the same job on the same commit SHA with no changes to any version-controlled file (code or CI configuration).}

\textbf{Industrial CI Context}. The present study has been conducted at TELUS, a major Canadian telecommunications company and our industrial partner. With over 35 years of existence, TELUS operates in 31 countries around the world, delivering cutting-edge IT solutions for connectivity. Notably, teams at TELUS are currently developing next-generation network softwarization products, such as SDNs \cite{noauthor_software-defined_nodate}. These SDNs are used for the dynamic and cost-effective management of large-scale network traffic in critical domains such as phone communications, internet access, television, connected agriculture, and telehealth \cite{telus_telus_2021}. In this context, delivering code frequently, safely, and \textbf{reliably} into production is paramount to achieving business objectives. Indeed, reliable delivery ensures minimal production issues, minimizes costly reworks and investigations, and improves developers' productivity.

Our industrial partner operates a highly distributed and heterogeneous technology ecosystem, including the CI/CD process used to integrate, validate, and deploy code changes. Central to this process is the self-hosted GitLab platform \cite{noauthor_gitlaborg_2024}, which serves as both version control and CI service. 
\added{The development teams at TELUS follow the GitFlow branching strategy~\cite{atlassian_gitflow_nodate}, a widely adopted model that defines five primary branch types. The \texttt{develop} branch serves as the central integration branch, where all ongoing work converges. Developers create \texttt{feature} branches from \texttt{develop} to implement new features or changes in isolation. Once validated, these feature branches are merged back into \texttt{develop}. Periodically, \texttt{release} branches are created from \texttt{develop} for final stabilization and testing before deployment, while \texttt{hotfix} branches address urgent production issues by branching directly from \texttt{main} (also referred to as \texttt{master}). The \texttt{main} branch reflects the latest stable production release. This branching model structures how code flows through the CI/CD pipeline and influences the types and frequencies of jobs triggered on each branch type.} 
\removed{GitLab, the most widely adopted self-hosted DevOps platform, owes its popularity to its support of the entire software development lifecycle and its open-core model, enabling organizations to securely host private instances. This approach is embraced by leading companies that have adopted GitLab (thus GitLab CI), such as IBM, Intel, SpaceX, Airbus, Siemens, Salesforce, and more than half of the Fortune 100 to securely manage their software codebases. Besides, the Software-as-a-Service (SaaS) version of GitLab (i.e., \texttt{gitlab.com}) hosts the second-largest collection of repositories after GitHub, especially since its acquisition by Microsoft. For these reasons, and given the need for contextual understanding of the projects, this study focuses on GitLab CI. . In our Online {Appendix} A, we provide additional details on GitLab CI as a leading CI platform, along with an extensive comparison to other CI services.}
\added{GitLab CI is the most widely adopted self-hosted DevOps platform~\cite{noauthor_23_nodate}, used by more than half of the Fortune~100~\cite{noauthor_gitlab_nodate}. Additional details on GitLab CI and its comparison to other CI services are provided in Online Appendix~A\footnote{\label{note:online_appendix}\url{https://figshare.com/s/0a7df599167838533bdb}}}.

In industrial CI environments, which are often self-hosted and distributed, unreliable \removed{build}\added{pipeline} outcomes frequently arise from irregular issues such as flaky tests or infrastructure problems \cite{lampel_when_2021, olewicki_towards_2022, aidasso_diagnosis_2025}. In the studied industrial setting, when a \removed{build}\added{pipeline} is triggered, GitLab CI assigns the defined jobs to self-managed runners, which execute them as soon as resources are available. In addition, most jobs require remote access to various resources such as code repositories, static code analysis servers, container platforms, and public cloud services. Given the organization’s strong security requirements, these accesses are protected by mechanisms like VPNs, SSL handshakes, and authentication tokens. As a result, \removed{build}\added{pipeline} jobs often fail irregularly due to factors unrelated to code changes  \cite{aidasso_diagnosis_2025}, such as sporadic unauthorized access, runner system faults, network instabilities, timeouts, and other unexpected errors.

\textbf{Silent Failures.} More concerningly, engineers at our industrial partner have reported cases where \removed{build}\added{pipeline} jobs are marked as successful despite failing to complete (or only partially completing) their specified tasks. We refer to this phenomenon as {silent failures}, which often go unnoticed for extended periods and erode developers' trust in the success of the CI/CD process. One such example is presented in this public issue\footnote{\label{note:security_scan_failure_ex1}\url{https://gitlab.com/gitlab-org/gitlab/-/issues/224187}} on GitLab, describing the case of a silent (license scanning) job failure, where...

\begin{fancyquote}
    ``\dots{the job succeeds, but actually fails silently without reporting anything [...] without any warning or error}\dots''
\end{fancyquote}

Another public issue\footnote{\label{note:artifact_failure}\url{https://gitlab.com/gitlab-org/gitlab/-/issues/22711}} reports successful jobs that fail to upload declared artifacts. The issue author underlined that...

\begin{fancyquote}
    ``\dots{As a result [of silent failures in successful jobs], one can introduce a regression in its CI/CD process without discovering it before days or client reporting incapacity to download artifact}\dots''
\end{fancyquote}

 Other examples noted during semi-structured interviews with five engineers at TELUS include container publication jobs that signal success even when no new image version is available in the registry, testing jobs that silently skip buggy files due to caching strategies, and many other similar issues. These silent failures lead to significant waste in the software development lifecycle in terms of delayed detection of issues or code defects, long manual verifications, delayed releases, and wasteful job reruns.

\textbf{Job Logs}. Each job produces extensive logs used by developers for debugging. These logs, generated from multiple commands and tools, are highly heterogeneous and can be very large, reaching up to thousands of lines and 50MB \cite{brandt_logchunks_2020}. Such volumes make diagnosing job failures difficult. But worst, silent failures typically show no clear signs in the logs\footref{note:artifact_failure}.

\begin{figure}%
    \centering
    \includegraphics[width=\linewidth]{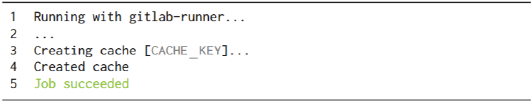}
\caption{Log excerpt of a silent cache upload failure in a successful job.}
\label{fig:silent_failure_log1}
\end{figure}

\begin{figure}
    \centering
    \includegraphics[width=\linewidth]{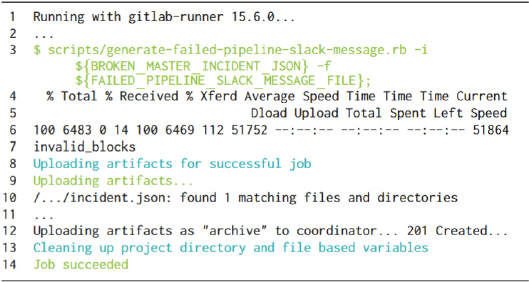}
\caption{Log excerpt of a silent HTTP request failure in a successful job.}
\label{fig:silent_failure_log2}
\end{figure}

\textbf{In many cases, it is very difficult (if not impossible) to identify silent failures in CI jobs only based on logs.} Fig.~\ref{fig:silent_failure_log1} shows an excerpt of execution logs of a successful job containing a cache upload failure as reported in this issue\footnote{\label{note:cache_error_example_1}\url{https://gitlab.com/gitlab-org/gitlab-runner/-/issues/2291}}. As can be observed, the logs contain no errors or warnings about the failure, although developers, as noted in the issue description, would have preferred to be notified of the problem.
 Another example is shown in Fig.~\ref{fig:silent_failure_log2}, from an issue\footnote{\url{https://gitlab.com/gitlab-org/gitlab/-/issues/384171}} on the GitLab project, where a \texttt{notify pipeline failure} job appears successful but silently fails to send pipeline incident report files to the teams. 
As shown in the figure, the logs appear to indicate full success of the job. On closer inspection, however, one can observe the small chunk \texttt{invalid\_blocks}, which reveals failure of the previous command. In large logs (e.g., exceeding 500 lines\footnote{\url{https://gitlab.com/gitlab-org/gitlab/-/jobs/3398283094}}), such a short and unclear error trace typically goes unnoticed, particularly by less experienced developers.

\textbf{Job Rerun Practice and Terminology.} When suspicious about a job outcome, developers tend to rerun the job multiple times without any change to the source code or \removed{build script}\added{CI configuration file} \cite{olewicki_towards_2022, aidasso_diagnosis_2025}, hoping for either (1) an actual job failure indicating the issues or (2) a true success with job tasks being actually completed \cite{hilton_usage_2016}. Fig.~\ref{fig:restart_suite} illustrates a job rerun scenario, where the first two successful executions of the \texttt{build and publish image} job are rerun, respectively leading to an additional execution of that job. For clarity, we define in the following the terminology used about job reruns in this paper.

Let $J_0^i, J_1^i, J_2^i, \dots, J_n^i$ denote the $n + 1$ consecutive executions of the same job $J$ on an identical code version, identified by the $i^{\text{th}}$ commit $c_i$ in a given project. Each subsequent job execution $J_{k+1}^i$ ($k \geq 0$) results from a manual rerun of the preceding execution $J_{k}^i$. For simplicity, we refer to each execution instance simply as a \textit{job}. A \textbf{job rerun sequence} is the set $\{J_0^i, J_1^i, J_2^i, \dots, J_n^i\}$ consisting of all (re)executions of a job $J$ on a given commit $c_i$. 
In each rerun sequence, we refer to the first job $J_0^i$ and the last job $J_n^i$ respectively as the \textbf{original} and \textbf{final jobs}. For each pair ($J_k^i \rightarrow J_{k+1}^i$) in the rerun sequence, we refer to $J_k^i$ as the \textbf{rerun-initiating job}, while $J_{k+1}^i$ denotes the \textbf{rerun job} resulting from manually rerunning $J_k^i$. Note that the original job is only a rerun-initiating job (and not a rerun job). Similarly, the final job is only a rerun job and not a rerun-initiating one. As a result, there is an equal number $n$ of rerun-initiating and rerun jobs in a job rerun sequence of $n + 1$ jobs. In an ideal wasteless CI/CD process, only the original job of a rerun sequence is executed. So, the additional rerun jobs in the sequence (i.e., all except the original job) are considered {wasteful} job executions. Finally, in the example of Fig.~\ref{fig:restart_suite}, silent job failures correspond to the first and second executions of the \texttt{build and publish image} job.

Although there is no official policy on rerunning suspicious jobs at TELUS, developers have confirmed the application of this practice in the event of silent failures. Therefore, in this paper, we study \textbf{silent job failures} by investigating jobs that were manually rerun (i.e., rerun-initiating jobs) despite having previously completed with a success signal. 
Repeatedly rerunning jobs increases the load on already constrained CI servers and leads to significant developer wait times, ultimately delaying software delivery \cite{olewicki_towards_2022, aidasso_diagnosis_2025}. As a result, silent job failures lead to significant waste for organizations, in particular TELUS, in terms of infrastructure and developer time.

Therefore, there is a strong need to better understand these silent job failures in order to support engineers in cutting their associated costs. Our findings will also provide valuable insights and guidelines that practitioners and CI providers can leverage to improve the reliability of CI/CD processes.

\section{Study Design}
\label{sec:study_design}

\begin{figure*}[tb]
    \centering
    \includegraphics[width=\textwidth]{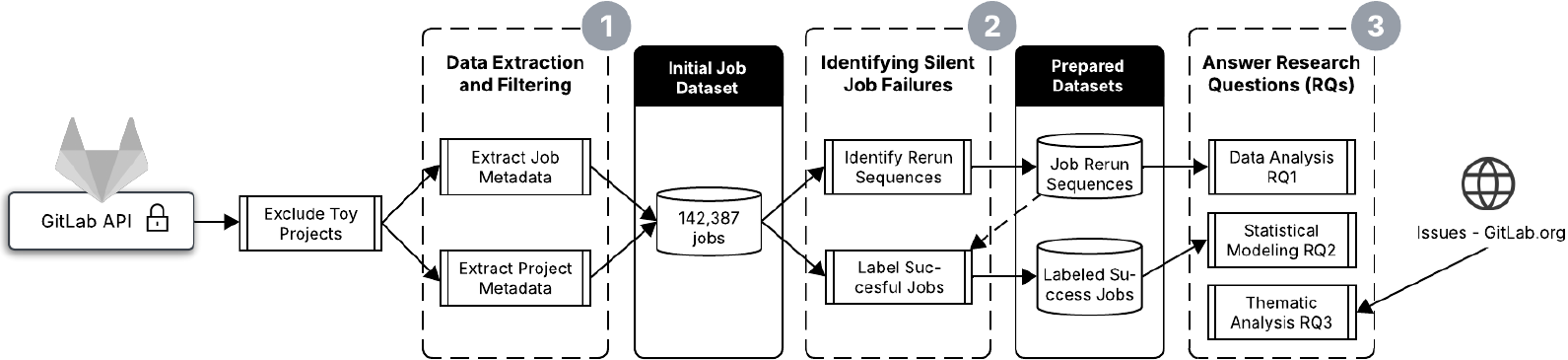}
    \caption{Overview of our approach for (1) data collection and (2) preparation, including data pre-processing and labeling, to (3) answer our research questions.}
\label{fig:study_design}
\end{figure*}

\subsection{Research Questions}

Our study aims to investigate the prevalence and impact of rerunning successful jobs due to silent failures (RQ1), identify the factors associated with rerunning successful jobs and thereby with silent failures (RQ2), and finally uncover the categories of silent failures that occur in successful jobs (RQ3) to better understand this phenomenon. For this purpose, we seek to answer the following research questions (RQs):

\begin{itemize}
    \item[] \rqone
    \item[] \rqtwo
    \item[] \rqthree
\end{itemize}

The motivations and approaches for answering each research question are further discussed in Section~\ref{sec:findings}. In the following, we present the main steps of our approach for data collection and preparation, illustrated in Fig.~\ref{fig:study_design}.

\subsection{Data Collection}
\label{sec:data_collection}

\textbf{Project Selection.} We identify with our industrial partner 14 main groups of actively maintained projects. From these groups, we filter out projects with fewer than 50 commits to exclude toy and experimental projects. As a result, we identified 81 microservice projects, from which we extracted the entire history of \removed{build}\added{pipeline} jobs up to July 11, 2024. Data collected from the oldest project span six years, while data from the youngest project span four months. These projects vary in size, purpose (e.g., backend APIs, network controllers, web portals, data analytics), and cover 13 programming languages, including Python, Go, C, TypeScript, JavaScript, and Shell. Our {Online Appendix B}\cref{note:online_appendix} lists the 81 selected projects and their characteristics, including number of commits, age, main programming language, purpose, and job execution statistics. Project names are omitted for privacy reasons.
 
\textbf{Data Extraction.} We collect raw metadata from \removed{build}\added{pipeline} jobs in the 81 identified projects, using the GitLab REST APIs\footnote{https://docs.gitlab.com/ee/api/jobs.html}. Specifically, we retrieve the raw metadata JSON files from a total of 153,374 jobs available in these projects. The JSON files are then flattened and consolidated into a CSV file. We only keep the data fields needed to compute the features listed in Table~\ref{tab:features}, which are discussed in detail in Section~\ref{sec:explanatory_factors}. Additionally, we collect project-level metadata, including the main programming language, project creation date, and the total number of commits.

\begin{table}
\caption{Overview of the Collected Data}
\begin{center}
\begin{tabular}{l r @{}}
\hline
\textbf{Metric} & \textbf{Value} \\
\hline
    \# Projects & 81 \\
    \# Languages & 13 \\
    \# Commits & 45,401 \\
    \# Pipelines & 53,851 \\
    \# \textbf{Jobs} & \textbf{153,374} \\
    \hline
    \# \textbf{Executed Jobs} & \textbf{142,387 (92.83\%)} \DrawPercentageBar{0.928} \\ %
    \# Success Jobs & 106,678 (69.55\%) \DrawPercentageBar{0.696} \\ 
    \# Failed Jobs & 29,983 (19.55\%) \DrawPercentageBar{0.195} \\
    \# Canceled Jobs & 5,726 (3.73\%) \DrawPercentageBar{0.0373} \\ %
    \# \textbf{Non-executed Jobs} & \textbf{10,987 (7.16\%)} \DrawPercentageBar{0.072} \\
    \# Skipped Jobs & 8,443 (5.50\%) \DrawPercentageBar{0.055} \\
    \# Manual Jobs & 2,450 (1.60\%) \DrawPercentageBar{0.016} \\
    \# Created Jobs & 94 (0.06\%) \DrawPercentageBar{0.00613} \\ \hline
    Date range of data collection & Feb 2018 --- July 2024 \\
\hline
\end{tabular}
\label{tab:collected_data}
\end{center}
\end{table}

Table~\ref{tab:collected_data} presents a summary of the collected data. We observe a ratio of 2.85 (i.e., 2 to 3) executed jobs per \removed{build}\added{pipeline} across the studied projects. Out of the 153,374 collected jobs, about 70\% completed successfully, underscoring the critical importance of ensuring the reliability of \textit{success} outcomes. The \textit{failed}, \textit{canceled}, \textit{skipped}, \textit{manual}, and \textit{created} jobs represent respectively 19.55\%, 3.73\%, 5.50\%, 1.60\%, and 0.06\%. 

\textbf{Data Filtering.} We filter out jobs with statuses \textit{skipped}, \textit{manual}, and \textit{created}, since these statuses indicate that the jobs were never executed \cite{gitlab_cicd_nodate} and thus provide no valuable information for the present study focusing on \removed{build}\added{pipeline} outcomes' reliability. Besides, these jobs only represent about 7\% of the collected data. We sort the remaining 142,387 executed jobs by date of creation (from the oldest to the most recent), forming the initial dataset of studied jobs.

\subsection{Identifying Silent Job Failures}
\label{sec:silent_failures_id}

\removed{In this study, rerun-initiating jobs having a \textit{success} status are considered as silent job failures.}
\added{\textbf{In this study, we use the rerun of successful jobs as a \textit{proxy} for silent job failures.} Rerun-initiating jobs having a \textit{success} status are therefore treated as candidate silent failures.}
 As discussed in Section~\ref{sec:background}, systematically identifying silent failures from the logs of successful jobs is impractical, as they often leave little to no noticeable trace of error (see Fig.~\ref{fig:silent_failure_log1} and Fig.~\ref{fig:silent_failure_log2}). Thus, we study silent job failures through the lens of successful jobs that developers have rerun. Indeed, based on interviews with TELUS engineers that confirmed the practice of rerunning suspected successful jobs, we hypothesize that successful jobs are rerun because they fail to complete their intended tasks, whether entirely or partially. \added{This approach follows the well-established practice of studying reruns of \emph{failed} jobs to characterize intermittent job failures~\cite{durieux_empirical_2020, olewicki_towards_2022, aidasso_diagnosis_2025}, which we extend to successful jobs. We validate this proxy through a manual analysis of 100 sampled jobs and discuss its limitations in Section~\ref{sec:threats}.} As illustrated in Fig.~\ref{fig:study_design}, we identify these jobs following the steps detailed below.

\textbf{Identifying Rerun Sequences.} We start by identifying jobs that were re-executed without any change to the source code or CI configuration file. To this end, we group the jobs by \textit{name}, \textit{commit} \added{ID}, and \textit{project}, in line with existing works \cite{olewicki_towards_2022, aidasso_efficient_2025, aidasso_diagnosis_2025}. \added{In GitLab CI, job names are unique within a project's pipeline configuration. This grouping therefore uniquely identifies a specific job definition executed multiple times on the same code version, with no change to any version-controlled file, including the CI configuration file (\texttt{.gitlab-ci.yml}).} From these groups, we identify job rerun sequences (see Fig.~\ref{fig:restart_suite}) as the job groups including more than one job, i.e., groups containing at least one additional execution of the same job on the same code version. As a result, we obtained a total of 13,380 job rerun sequences, each containing jobs in chronological order of execution. 
We use the dataset of job rerun sequences to answer our RQs. Specific approaches are detailed in Section~\ref{sec:findings}.

\textbf{Labeling Successful Jobs.} To address RQ2, which investigates the factors distinguishing rerun successful jobs (silent failures) from non-rerun successful jobs (true successes), we focus on the dataset of 106,678 successful jobs. We label this dataset using the 13,380 job rerun sequences by identifying the IDs of rerun-initiating jobs having a \texttt{success} status. 
\added{This filtering on the \texttt{success} status ensures that jobs that were accidentally canceled and subsequently rerun are excluded from the RQ2 analysis by design.}
The successful jobs matching these IDs in the dataset are labeled as having been rerun (\texttt{rerun = 1}), while all others are labeled as non-rerun (\texttt{rerun = 0}). As such, the variable \texttt{rerun} indicates whether a successful job was rerun or not and is the dependent variable that we seek to explain in RQ2. The results of this labeling process, which shed light on the prevalence of silent failures in re-executed successful jobs, are discussed in RQ1.

\begin{table}[]
\noindent\begin{minipage}{\linewidth}
\centering
\caption{Description and Rationale of Factors used as Independent Variables. ``F." stands for Family.}\label{tab:features}
\begin{tblr}{
  width = \linewidth,
  colspec = {Q[30]Q[892]},
  cell{2}{1} = {r=3}{},
  cell{5}{1} = {r=6}{},
  cell{11}{1} = {r=3}{},
  vline{2} = {1-15}{},
  hline{1-2,5,11,14} = {-}{},
  hline{3-4,6-10,12-13} = {2}{},
}
\textbf{F.}                                             & \textbf{Description/Rationale}                                                                                                                                                                                                                                                                                                                           \\
\begin{sideways}Job Execution Context\end{sideways}     & {\textbf{job\_type}: Type of executed job derived from the job name (e.g., static analysis, testing, deployment).\\\uline{Rationale}: Inspired by \cite{rausch_empirical_2017}, reruns of successful jobs might be associated with some specific types of jobs (e.g., deployment jobs).}                                                                                             \\
                                                        & {\textbf{branch\_type}:\textbf{ }The type of branch on which the job was triggered (e.g., main, develop, feature, hotfix).\\\uline{Rationale}: Reruns of successful jobs might be associated with jobs triggered on specific branches, since the stability expectations are different across branches \cite{islam_insights_2017}.}            \\
                                                        & {\textbf{month\_of\_year}, \textbf{day\_of\_week}, \textbf{hour\_of\_day}: The moment when the job is triggered. Values range in [1–12], [1-7], and [0-23].\\\uline{Rationale}: Inspired by this issue \cite{noauthor_docker_2019} and prior work \cite{weeraddana_characterizing_2024, kola-olawuyi_impact_2024, ghaleb_empirical_2019}, different load periods on the CI server can explain success reruns.}                                                                     \\
\begin{sideways}Project (History) Factors\end{sideways} & {\textbf{language}: The main programming language of the project.\\\uline{Rationale}: Successful job reruns might be more or less common in projects written in specific programming languages \cite{ghaleb_empirical_2019, benjamin_study_2023}.}                                                                                                                                                         \\
                                                        & {\textbf{age:} Age of the project in days when the job is triggered.\\\uline{Rationale}: Inspired by \cite{benjamin_study_2023, kola-olawuyi_impact_2024}, older or younger projects may be more or less prone to instabilities and reruns of successful jobs.}                                                                                                                                              \\
                                                        & {\textbf{commits}: Number of commits at the time the job is triggered.\\\uline{Rationale}: Mature or complex projects may more easily include silent regressions explaining successful job reruns \cite{benjamin_study_2023}.}                                                                                                                             \\
                                                        & {\textbf{\textbf{build\_frequency:}} Average \#jobs per day when the job is triggered.\\\uline{Rationale}: A high job frequency may stress CI resources, increasing the likelihood of silent issues and of rerunning successful jobs.}                                                                              \\
                                                        & {\textbf{\textbf{failure\_ratio}}: Percentage of failed jobs when the job is triggered.\\\uline{Rationale}: Projects with lower failure rates might still suffer from silent issues and thus prompt to rerunning successful jobs \cite{kola-olawuyi_impact_2024}.}                                                                                     \\
                                                        & {\textbf{\textbf{\textbf{\textbf{rerun\_ratio:}}}} Percentage of rerun jobs, when the job is triggered.\\\uline{Rationale}: Inspired by \cite{olewicki_towards_2022}, a high proportion of prior rerun jobs may suggest reliability issues within the project, increasing the likelihood of rerunning successful jobs.}                                        \\
\begin{sideways}Developer Factors\end{sideways}         & {\textbf{author\_exp\_days:}~Experience (number of days since first contribution) of the developer who authored the triggering commit.\\\uline{Rationale}:~Less experienced developers may have a less thorough understanding of projects \cite{ghaleb_empirical_2019, rausch_empirical_2017} and are more likely to rerun successful jobs when unsure of the outcome.}                 \\
                                                        
                                                        & {\textbf{\textbf{author\_exp\_commits:~}}Experience (ratio of total commits in the project) of the developer who authored the triggering commit.\\\uline{Rationale}: Inspired by \cite{ghaleb_empirical_2019, rausch_empirical_2017, kola-olawuyi_impact_2024}, the ratio of commits is another estimation of the author's experience, which, similarly to author\_exp\_days, might explain rerun of successful jobs.} \\
                                                        
                                                        & {\textbf{\textbf{author\_recent\_rerun\_history: }}Proportion of rerun jobs among the last five jobs triggered by the author of the current job.\\\uline{Rationale}: Inspired by \cite{weeraddana_characterizing_2024, kola-olawuyi_impact_2024}, the author tendency to rerun jobs suggests they might be working on change sets associated with reliability issues, increasing the odds of rerunning successful jobs.}                                                                                     
\end{tblr}
\end{minipage}
\end{table}

\subsection{Independent Explanatory Variables}
\label{sec:explanatory_factors}

Similar to prior studies \cite{weeraddana_characterizing_2024, benjamin_study_2023, hong_practitioners_2024}, we identify a set of independent variables aimed at uncovering the most influential factors associated with the rerun of successful jobs, i.e. of silent job failures (RQ2). Table~\ref{tab:features} presents the initial list of 14 variables with their descriptions and the rationale for studying them. These variables are grouped into three families. First, the \textit{job execution context} family includes job-level factors potentially associated with silent issues that prompt developers to rerun successful jobs. Inspired by previous studies \cite{benjamin_study_2023, ghaleb_empirical_2019, kola-olawuyi_impact_2024}, the other two variable families, \textit{project factors} and \textit{developer factors}, enable us to assess how project characteristics and developer profiles relate to the rerun of successful jobs. The insights from these associations can help managers identify and focus efforts on the most critical types of projects and help raise awareness among teams.

\added{The independent variables used in our model draw on feature families commonly used across CI factor modeling studies, where they have been applied to different dependent variables such as build duration~\cite{ghaleb_empirical_2019}, build timeouts~\cite{weeraddana_characterizing_2024}, build failures~\cite{rausch_empirical_2017, islam_insights_2017}, and recheck outcomes~\cite{brus_rechecking_2025}. In our study, the dependent variable is whether a \emph{successful} job is rerun, a proxy for silent failure. Reusing established feature families to study a new outcome enables us to build on factors whose practical relevance has been demonstrated in prior CI research.}

In our {Online Appendix C},\cref{note:online_appendix} we describe the feature engineering process for the categorical independent variables (i.e., job\_type, branch\_type, and language) and show the distribution of categories for each of these variables. \added{In particular, we assigned branch types based on the GitFlow naming convention~\cite{atlassian_gitflow_nodate} adopted in the studied projects, which includes five primary types: \texttt{main}, \texttt{develop}, \texttt{feature}, \texttt{release}, and \texttt{hotfix} (see Section~\ref{sec:background} for details).} In particular, we identified 7 job types (testing, deployment\_and\_release, static\_code\_analysis, image\_security\_scan, code\_and\_image\_build, notification, and other) via open-coding \cite{corbin_sage_nodate} and label consolidating, 5 branch types (develop, feature, release, hotfix, and main) based on the GitFlow naming convention, and 10 main languages (Python, Gherkin, Go, HCL\footnote{HCL is a domain-specific language for defining Infrastructure-as-Code}, HTML, JS, Jinja, Shell, TSX, and TS). The remaining numerical independent variables are directly computed from the collected data fields as described in Table~\ref{tab:features}.

\subsection{One-Hot Encoding of Categorical Variables}
\label{sec:one_hot_encoding}

We use one-hot encoding \cite{noauthor_onehotencoder_nodate} to transform the categorical variables (job\_type, branch\_type, and language) into binary dummy variables  \cite{kumar_logistic_2021}. For each variable, one-hot encoding creates multiple dummies, one for each of its category levels. A dummy variable having a value of 1 indicates that the observation (i.e., the job) belongs to the corresponding category level, while 0 indicates it does not. As a result, we derived 7 dummy variables for job types (e.g., job\_type\_testing), 5 for branch types (e.g., branch\_type\_feature), and 10 for the language variable (e.g., lang\_python). In statistical modeling, one dummy is typically removed for each categorical variable to prevent multicollinearity and serve as a reference category for interpreting the remaining dummies' log-odd coefficients \cite{kumar_logistic_2021, ghaleb_empirical_2019}. The choice of the reference to drop is described further in Section~\ref{sec:rq2}. At this stage, we obtained 33 numerical (independent) variables and one dependent binary variable for the 106,678 successful jobs in our dataset for RQ2.

\subsection{Cyclical Encoding of Periodic Variables}
\label{sec:cyclical_encoding}

A subset of our variables captures time periods, i.e., hour\_of\_day, day\_of\_week, and month\_of\_year, with 23, 7, and 12 category levels, respectively. 
However, treating these variables as categorical like in \cite{ghaleb_empirical_2019} would (1) require creating 42 additional dummy variables (23 + 7 + 12), increasing the complexity of the model, and (2) imply creating artificial discontinuities at period boundaries (e.g., between hour 23 and 0). In fact, one-hot encoding ignores the natural cyclical patterns of temporal variables, which could nevertheless offer more meaningful insights across broader periods (e.g., end of year) rather than fixed categories like individual months or days. To capture the cyclical patterns in these variables, we apply cyclical encoding \cite{bansal_temporal_2025}, which transforms each variable $t$ into two new sinusoidal components as follows:
\[
\left\{
\begin{array}{l}
x_{\sin}(t) = \sin\left(2\pi \cdot \frac{t}{P}\right) \\
x_{\cos}(t) = \cos\left(2\pi \cdot \frac{t}{P}\right)
\end{array}
\right.
\quad \text{where} ~ P = \max(t)
\]

As a result, each temporal variable is encoded into sine and cosine variables representing their (x, y) coordinates on a circle, where period boundaries (e.g., hour 23 and 0, or December and January) are closer. Therefore, we derive six (2 $*$ 3) new variables in place of the three temporal variables, resulting in a total of 36 independent variables. 

\begin{tcolorbox}[rqbox={Summary of Datasets}]
We use the 13,380 job rerun sequences (36,489 jobs) for RQ1, and the 106,678 successful jobs across 81 projects for RQ2, each with 36 numerical independent variables and one dependent variable for statistical modeling.
\end{tcolorbox}

\section{Study Results}
\label{sec:findings}

\subsection{\textbf{\rqone}}
\label{sec:rq1}

\textbf{Motivation.} The goal of this first RQ is to evaluate the prevalence and cost of the practice of rerunning jobs at TELUS, especially jobs that had already been completed successfully. Indeed, we leverage reruns of successful \removed{build}\added{pipeline} jobs as an indicator of silent failures, based on developers' experience with silent failures as reported during interviews. However, since there is no explicit policy at TELUS recommending reruns in response to unreliable outcomes or silent failures, we first seek to examine the prevalence and impact of such practice. The findings from this RQ will offer valuable insights into the practice of rerunning successful jobs, laying the groundwork and motivation for a deeper investigation of silent failures and associated job reruns.

\textbf{Approach.} To answer RQ\textsubscript{1}, we identify jobs that were rerun without any modifications to the source code or \removed{build script}\added{CI configuration file}. Following the method described in Section~\ref{sec:silent_failures_id}, we identify the job rerun sequences (illustrated in Fig.~\ref{fig:restart_suite}), resulting in a dataset of 13,380 sequences. In a rerun sequence containing $n$ jobs, $n - 1$ are rerun jobs, i.e., all but the original job. Similarly, $n -1$ are rerun-initiating jobs, i.e., all but the final job. Using the dataset of job rerun sequences, we compute statistics on rerun jobs (overall and by rerun-initiating job status) and analyze the distribution of rerun counts (i.e., rerun sequence lengths minus one) across projects. In addition, similarly to prior work \cite{durieux_empirical_2020}, we investigate the distribution of status change in the pairs of rerun-initiating jobs and subsequent rerun jobs to better assess how often developers rerun successful jobs versus failed or canceled ones. Furthermore, we analyze the distribution of time gaps between the start of each rerun-initiating job and its subsequent rerun within the sequences, by rerun-initiating job status, in order to assess the delays before noticing (silent vs. actual) failures and diagnosing them. Finally, we evaluate the amount of machine time spent running wasteful rerun jobs that would be saved if only the original jobs had been executed in each rerun sequence.

\begin{table}
\caption{Statistics of Job Reruns at Telus}
\begin{center}
\begin{tabular}{l r @{}}
\hline
\textbf{Metric} & \textbf{Value} \\
\hline
    \# Projects including rerun jobs & 81 (100\%) \DrawPercentageBar{1} \\ 
    \# Rerun jobs & 23,109 (16.23\%) \DrawPercentageBar{0.162} \\ 
    \# Rerun sequences & 13,380 \\
    \hline
    \# Sequence of $1$ rerun & 9,317 (69.63\%) \DrawPercentageBar{0.696} \\
    \# Sequence of $2$ reruns & 2,263 (16.91\%) \DrawPercentageBar{0.169} \\
    \# Sequence of $3$ reruns & 865 (6.46\%) \DrawPercentageBar{0.065} \\
    \# Sequence of $4$ reruns & 379 (2.83\%) \DrawPercentageBar{0.028} \\
    \# Sequence of $5$ reruns & 186 (1.39\%) \DrawPercentageBar{0.014} \\
    \# Sequence of $\geq 6$ reruns & 370 (2.76\%) \DrawPercentageBar{0.028} \\
    Avg \# of jobs per sequence & $\approx$ 2 \\
    Max \# of reruns for a single job  & 29 
    \\
\hline
\end{tabular}
\label{tab:reruns_statistics}
\end{center}
\end{table}

\begin{figure}
  \begin{center}
      \includegraphics[width=.8\linewidth]{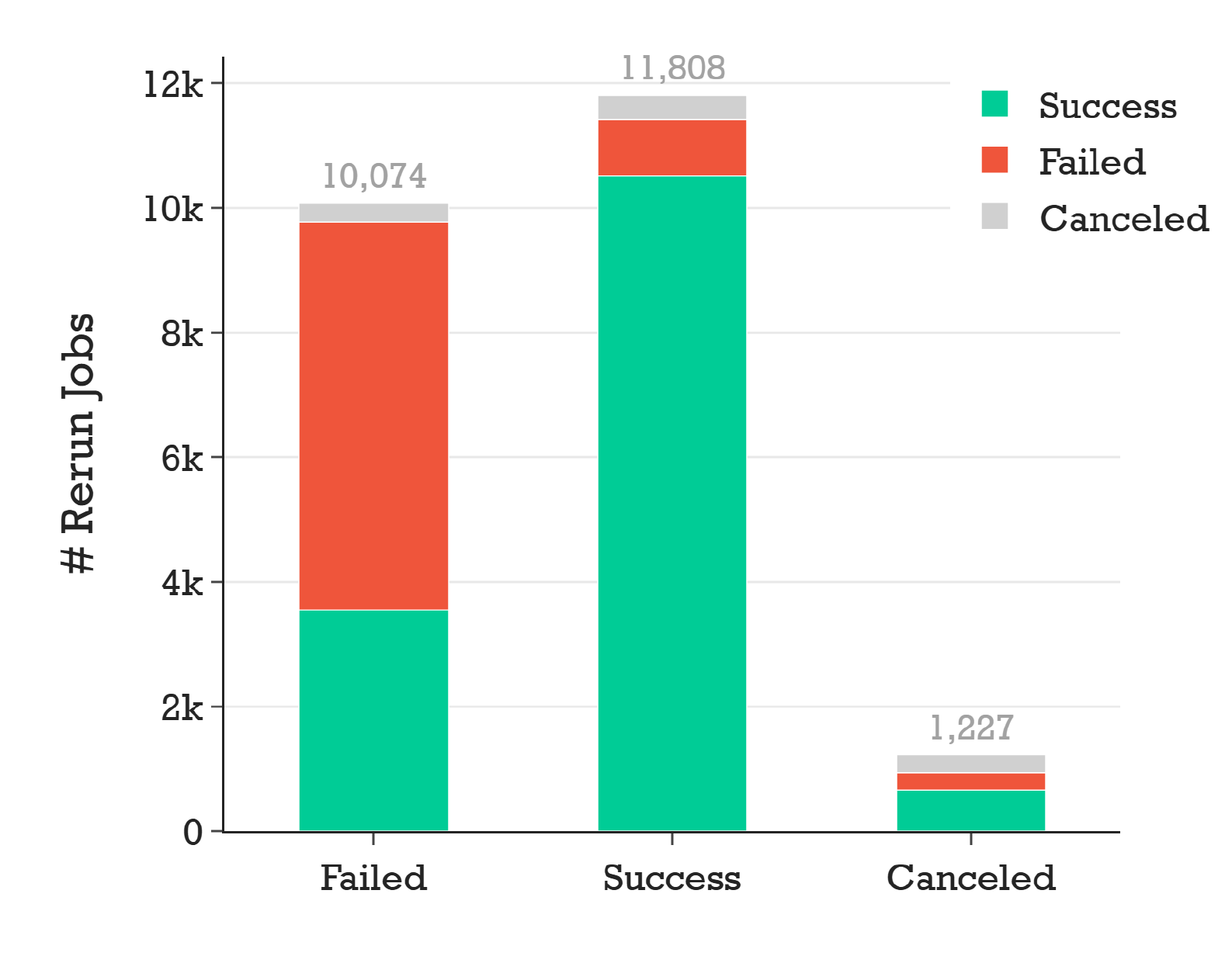}
  \end{center}
\caption{Evolution of job statuses between rerun-initiating jobs (x-axis) and the immediate rerun jobs (y-axis) in the identified job rerun sequences.}
\label{fig:status_evolution}
\end{figure}

\textbf{Results.} Table~\ref{tab:reruns_statistics} presents job rerun statistics for the 81 studied projects, including the distribution of rerun counts and the number of reruns by rerun-initiating job status. Fig.~\ref{fig:status_evolution} shows the distribution of status transitions between each job and its immediate rerun in the sequences.

\textbf{A significant proportion (approximately 16\%) of the studied jobs have been rerun, highlighting the prevalence of this practice. On average, these reruns occur in sequences involving two additional runs of the same job, with the number of reruns reaching up to 29.}  As reported in Table~\ref{tab:reruns_statistics}, we found 23,109 rerun jobs, accounting for 16.23\% of all 142,387 studied jobs (including passed, failed, and canceled ones). The rerun ratio (16.23\%) is strikingly high in the industrial setting, almost ten times that (1.7\%) reported in a prior open-source study \cite{durieux_empirical_2020} and more than double the ratio (7.95\%) observed in our replication study on open-source projects detailed in Online Appendix F\cref{note:online_appendix}. \added{We note that direct comparison of rerun ratios across studies requires caution, as they operate at 
different granularities and in different ecosystems. Durieux et al.~\cite{durieux_empirical_2020} report 1.72\% of restarted \emph{builds} on Travis~CI (pipeline level), while our ratio of 16.23\% measures reruns at the \emph{job} level within a pipeline, which is a finer granularity. A single pipeline typically contains multiple jobs, making these ratios not directly comparable.}
Besides, rerun jobs were found in all 81 studied projects, further emphasizing the widespread nature of this practice across the industrial projects. The 23,109 rerun jobs are found in a total of 13,380 rerun sequences, averaging 1.73 reruns per sequence (i.e., typically one to two reruns). Specifically, the majority of rerun sequences (69.63\% $\approx$ 70\%) consist of a single rerun. The remaining 17\%, 6\%, 3\%, 1\%, and 3\% of job rerun sequences involve 2, 3, 4, 5, and more than 5 (up to 29) re-executions of the same job, respectively. 

\textbf{More than half (51\%) of the rerun-initiating jobs were successful, while failed ones account for 44\%. This result indicates a significant presence of silent job failures ($\approx$ 11\%) among all jobs marked as successful, and emphasizes the need for further investigation to understand their associated factors and impact.} As previously described, 23,109 rerun jobs imply 23,109 rerun-initiating jobs identified in the rerun sequences. Of these 23,109 rerun-initiating jobs, 11,808 (i.e., 51.08\% or more than half of the rerun-initiating jobs) are jobs that have initially completed their execution with a \textit{success} signal. These jobs represent 11\% of all the studied successful jobs. In contrast, failed jobs represent 44\% of the rerun-initiating jobs, while canceled jobs account for the remaining 5\%.
The notably high proportion of successful jobs in the rerun-initiating ones, surpassing that of failed jobs, is particularly intriguing. Intuitively, one would expect failed jobs to be the primary trigger for reruns, as they more clearly warrant developer intervention. Nevertheless, this finding aligns with developers’ reports during interviews, where they described frequently encountering silent failures hidden under signals of job success. It also emphasizes the need to investigate the factors associated with the reruns of successful jobs, in order to better understand this underexplored phenomenon.

\begin{figure}
  \begin{center}
      \includegraphics[width=\linewidth]{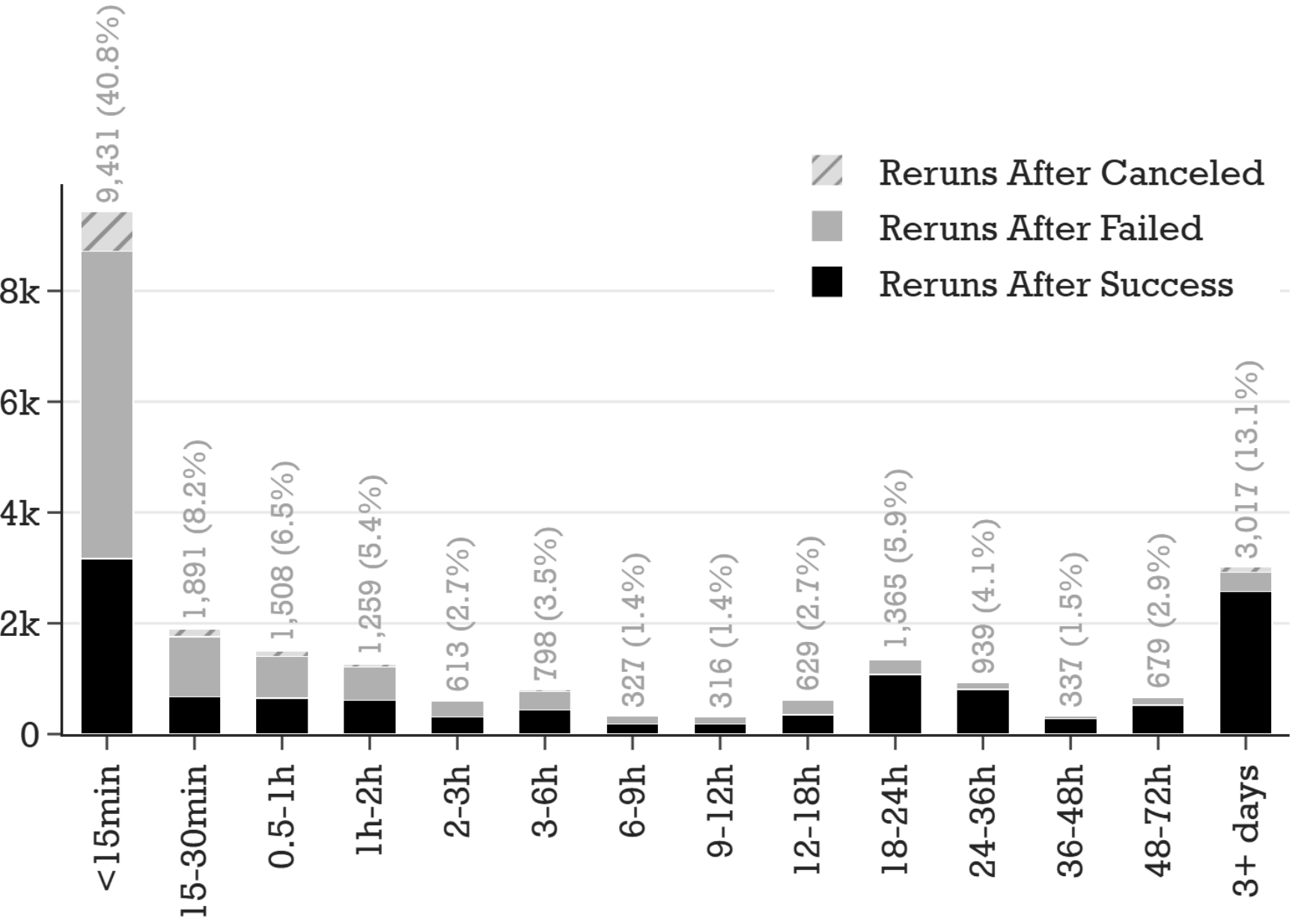}
  \end{center}
\caption{Distribution of time delays between rerun-initiating jobs and job reruns, by status of the initiating jobs. Each bar is annotated with the number of rerun jobs in that interval, with the corresponding proportion in parentheses.}
\label{fig:time_gap_dist}
\end{figure}

\textbf{
Reruns after successful jobs happen much later (median 6.9 hours, 4.6 days on average) compared to reruns after failed jobs (median 11.4 minutes, 13.3 hours on average) and canceled jobs (median 8.4 minutes, 1.3 days on average). This shows that silent failures often go unnoticed for long periods and suggests they require longer diagnostic times.} Fig.~\ref{fig:time_gap_dist} shows the distribution of delays between the creation times of each rerun-initiating job and its immediate rerun. It highlights in black the distribution of time delays for instances where the rerun-initiating jobs were successful, i.e. delays before rerunning successful jobs. As shown in the figure, a substantial portion (40.8\%) of reruns takes place within 15 minutes, regardless of the rerun-initiating job status. Notably, over half of all reruns (55.5\%) occur within one hour, and the vast majority (73.3\%) of these reruns were initiated following failed jobs. These results indicate that developers respond quickly to explicit failures. In contrast, reruns following successful jobs follow two different patterns, as can be seen in Fig.~\ref{fig:time_gap_dist}. Indeed, 31.4\% of these reruns are initiated within 15 minutes, while a further 30.6\% are initiated only after 2 days. Moreover, the majority (52.03\%) of reruns following successful jobs take place only after a one-day delay. Discussions with developers at TELUS suggest that this is because, on the one hand, some silent job failures (e.g., deployed container not running in production) are noticed quite quickly when developers already suspect the job and are actively awaiting its effects via manual verifications in production, for instance. In these cases, reruns are performed immediately in the hope that the job will actually complete its task. Alternatively, the silent failures are either only noticed very late on or require the intervention of infrastructure teams to diagnose them, thus extending the time delay for re-executing the job to ensure that it finally works as intended.

\textbf{Rerun jobs following success account for nearly half (48\%) of the total wasted server time from all rerun jobs. This finding emphasizes that silent failures significantly contribute to increased CI costs.} The median duration of rerun jobs following a successful execution (1.75 minutes) is higher than that of rerun jobs after a failure (1.5 minutes). As a result, rerun jobs after success spent a substantial amount of server time, i.e. 29 days and 13 hours, which accounts for 47.5\% of the total server time of 62 days and 6 hours for all reruns. \added{To put this in perspective, the total CI compute time across all 142,387 studied jobs is approximately 359 days. The 62 days spent on all reruns therefore represent 17.3\% of the total CI compute budget, and the 29.5 days spent on reruns of successful jobs alone account for 8.2\%.}
Thus, much of the wasteful server usage due to reruns, avoidable if the original executions had sufficed, stems from silent failures that prompt the rerun of successful jobs.

\added{\textbf{Replication of RQ\textsubscript{1} on OSS projects.} To assess the generalizability of our findings, we conducted a replication study on 198,146 jobs across 8 popular open-source GitLab projects (details in Online Appendix~F\cref{note:online_appendix}). We found that although the overall rerun ratio is lower in OSS projects (7.95\% vs.\ 16.23\%), reruns of successful jobs are even more predominant, accounting for approximately 76\% of all job reruns (vs.\ 51\% in the TELUS industrial setting). Furthermore, similar to the industrial context, reruns of successful jobs account for almost all reruns occurring after 6~hours. These results provide initial evidence that the practice of rerunning successful jobs extends beyond the studied industrial context.}

\begin{tcolorbox}[rqbox={Summary of RQ\textsubscript{1}}]
About 11\% of successful jobs are rerun, accounting for more than half (51\%) of all rerun-initiating jobs. The reruns occurred after longer time delays than those of failed and canceled jobs, and also consumed 48\% of the total server time spent on all reruns. These results highlight the prevalence and stealthy nature of silent failures (causing late detection) and their high contribution to CI costs.
\end{tcolorbox}

 \subsection{\textbf{\rqtwo}}
\label{sec:rq2}

\textbf{Motivation.} We identified in RQ\textsubscript{1} a significant number of initially successful jobs that have been later rerun. Hence, the goal of this RQ is to investigate whether such reruns can be explained by project and job characteristics. If so, we further seek to identify the key factors that most strongly influence the rerun of successful jobs. These findings will provide practitioners with valuable insights into the circumstances surrounding silent failures and help raise awareness among teams to support their effective identification and mitigation.

\textbf{Approach.} To identify the most important factors associated with the rerun of successful jobs, we model the likelihood of developers rerunning successful jobs based on various independent variables listed in Table~\ref{tab:features}. For this purpose, we use mixed-effects logistic regression models\footnote{\url{https://stats.oarc.ucla.edu/r/dae/mixed-effects-logistic-regression/}}, a type of generalized linear mixed-effects model (GLMM), that emphasizes interpretability and enables us to capture both fixed and random effects across groups (i.e., projects in our case) \cite{faraway_extending_2016}. 

\begin{equation}
    \label{eq:logit_fn}
    \boxed{
        ~~~\texttt{{logit}}(P(Y_g = 1)) = \beta_0 + \sum_{i=1}^{n} \beta_i X_i + \theta_g~~~
    }
\end{equation}
\vspace{1pt}

Eq.~\ref{eq:logit_fn} shows the equation of a mixed-effects logistic regression model. It estimates the log-odds of the binary outcome \( Y_{g} \) as a linear combination of two types of effects: fixed effects \( \boldsymbol{\beta_i} \) and random effects \( \mathbf{\theta}_g \). The fixed effects (\( \boldsymbol{\beta_i} \)), constant across all groups, are the coefficients of the independent variables \(X_i\), while the random effects (\( \mathbf{\theta}_g \)) capture group-specific variations \cite{faraway_extending_2016, ghaleb_empirical_2019}, such as differences across projects. \(\beta_0\) denotes the constant intercept representing the baseline log-odds of the outcome when all predictors are zero.

In line with prior work \cite{weeraddana_characterizing_2024, ghaleb_studying_2023, ghaleb_studying_2019}, the choice of a statistical inference model (over traditional machine learning (ML) models) is driven by our primary objective of explanation rather than prediction. Indeed, statistical models provide interpretable coefficients, allowing us to quantify the influence of each variable on the predicted outcome, unlike ML models, which typically operate as black boxes\footnote{\url{https://www.fharrell.com/post/stat-ml}}.

The initial input dataset of our approach consists of the 106,678 successful jobs, including 36 independent variables and one dependent variable (i.e., \texttt{rerun}), as detailed in Sections~\ref{sec:silent_failures_id} and \ref{sec:explanatory_factors}. Our dataset is imbalanced with only 11\% of positive examples, as discussed in RQ\textsubscript{1}. While class balancing techniques such as SMOTE \cite{chawla_smote_2002} are commonly used in ML to improve the recall of the minority class, according to the recent findings of Weeraddana et al. \cite{weeraddana_characterizing_2024}, this technique does not lead to substantial differences in the outcomes of statistical models. Therefore, we do not apply class balancing to our dataset, which also enables our model to more accurately reflect real-world scenarios in line with \cite{weeraddana_characterizing_2024}. 
\added{While additional sensitivity analyses such as downsampling the majority class could be explored, our model's EPV ratio of 369 far exceeds the recommended threshold of 10~\cite{peduzzi_simulation_1996}, and is well-calibrated as discussed in the results. These indicators provide strong evidence against overfitting and suggest that class imbalance is unlikely to affect our results.}
To create the mixed-effects model, we follow the guidelines for statistical regression modeling provided by Harrell \cite{harrell__regression_2015}, which we describe in the five steps below.

\textbf{{Step 1. Correlation Analysis.}} Highly correlated variables are known to distort statistical models \cite{weeraddana_characterizing_2024, ghaleb_studying_2023}, undermining the interpretation of contributions of each independent variable to the dependent variable.
Therefore, we sought to address high correlations among variables using Spearman’s rank correlation ($\rho$), which (in contrast to Pearson's) captures both linear and non-linear relationships \cite{weeraddana_characterizing_2024}. For this purpose, we use the \texttt{varclus} function from the \texttt{Hmisc}\footnote{\label{note:hmisc}\url{https://cran.r-project.org/web/packages/Hmisc}} package in R. In line with prior work \cite{weeraddana_characterizing_2024}, we use a threshold of $|\rho| = 0.7$ to determine pairs of highly correlated variables. As a result, we identify one pair of highly correlated variables (\texttt{age}, \texttt{commits}), from which we remove one variable to mitigate the collinearity \cite{weeraddana_characterizing_2024, ghaleb_empirical_2019, kola-olawuyi_impact_2024, kumar_logistic_2021}.
Specifically, we remove the variable \texttt{age}, as \texttt{commits} offers a more meaningful estimate of project maturity and complexity. 
Complete results of our correlation analysis are provided in our {Online Appendix D}.\cref{note:online_appendix}

\textbf{{Step 2. Redundancy Analysis.}} We perform a multicollinearity analysis on the 35 remaining independent variables to identify and mitigate redundant features. Multicollinearity occurs when a feature can be linearly explained by the combination of other features (e.g., a dummy variable that can be derived from the others \cite{kumar_logistic_2021}). For this purpose, we use the \texttt{redun} function from the \texttt{Hmisc}\cref{note:hmisc} package in R, which fits 35 separate models, each predicting one feature based on the remaining 34. Similar to prior work \cite{weeraddana_characterizing_2024, ghaleb_empirical_2019}, we use an $R^2$ threshold of 0.9 to identify features recommended for exclusion due to redundancy. As a result, we identify and remove the redundant variables: \texttt{\small branch\_type\_main}, \texttt{\small lang\_python}, and \texttt{\small job\_type\_deployment\_and\_release}. In fact, these variables are dummies for the original variables \texttt{\small branch\_type}, \texttt{\small lang}, and \texttt{\small job\_type}, respectively, and therefore serve as reference categories for these variables. At the end of the redundancy analysis, our dataset includes 32 independent variables, which we use in our model.

\textbf{{Step 3. Mixed Effects Model Fitting.}} We fit our mixed-effects logistic regression model using the \texttt{glmer} function from the \texttt{lme4}\footnote{\url{https://cran.r-project.org/web/packages/lme4}} package in R. It models the binary outcome \texttt{rerun} using the 32 independent numerical variables as fixed effects and the job's project ID as the grouping factor for random effects. 
\added{Our model uses a random intercept-only structure, consistent with prior mixed-effects models in CI research~\cite{weeraddana_characterizing_2024, ghaleb_empirical_2019}. Adding random slopes across 81 projects with 32 fixed effects would substantially increase model complexity and risk convergence issues. Moreover, the marginal $R^2$ obtained for our random intercept-only model already accounts for 83\% of the conditional $R^2$, suggesting that fixed effects capture most of the explained variance. We leave the exploration of random slopes to future work.}
Following Ghaleb et al. \cite{ghaleb_empirical_2019}, we specify the \textit{binomial} distribution, \textit{Laplace} approximation, and \textit{bobyqa} optimizer as parameters of the \texttt{glmer} function. Furthermore, we calculate the Events Per Variable (EPV) ratio to assess the risk of overfitting in our model \cite{peduzzi_simulation_1996}. An EPV above 10 indicates a low risk of overfitting in logistic models \cite{peduzzi_simulation_1996, ghaleb_empirical_2019}. 

\textbf{{Step 4. Performance Evaluation.}} \removed{In line with prior work, we evaluate our model according to (a) its discriminatory power, (b) its ability to balance precision and recall, (c) the calibration of its predicted risk estimates, and (d) its overall goodness-of-fit. For this purpose, we use the following metrics:} \added{We evaluate our model using five standard metrics: ROC AUC~\cite{hanley_meaning_1982}, AUC-PR~\cite{saito_precision-recall_2015}, Brier Score~\cite{brier_verification_1950}, Marginal $R^2$~\cite{mcfadden_quantitative_1978}, and Conditional $R^2$. Definitions and interpretation guidelines for each metric are provided in Online Appendix.\cref{note:online_appendix}}

\removed{\textit{Detailed list of metrics.}}

\textbf{{Step 5. Significant Factors Identification.}} We estimate the importance of each independent variable using the maximum likelihood $\chi^2$ (Chi-Squared) test, also known as the likelihood ratio $\chi^2$ test, used in prior studies \cite{weeraddana_characterizing_2024, ghaleb_empirical_2019}. An independent variable is significant if it has $Pr(< \chi^2) < 0.05$, where $Pr(< \chi^2)$ is the p-value from the $\chi^2$ statistical test. Its associated $\chi^2$ value indicates whether the model is statistically different from the same model in the absence of the independent variable. A larger $\chi^2$ value indicates that the independent variable has a stronger explanatory power of the rerun of successful jobs.

\textbf{Results.} We report the model performance in Table~\ref{tab:model_performance}. Table~\ref{tab:feature_importances}  shows, for each independent variable, the model coefficient (i.e., the log-odds), the $\chi^2$ test results, and the odds ratio ($\text{OR} = e^{log~odds}$). Significant variables are marked with asterisks. Positive or negative log-odds indicate that the variable has a direct relationship (i.e., an increase in the variable increases the likelihood of rerunning successful jobs) or an inverse relationship, respectively. In the following, we discuss the performance of our model, and for synthesis, the most significant and practically relevant factors that are identified. Statistical interpretations of all significant factors are provided in our {Online Appendix E}.\cref{note:online_appendix}

\begin{table}
\caption{Model Performance}
\label{tab:model_performance}
\centering
\begin{tblr}{
  width = .82\linewidth,
  colspec = {Q[250]Q[210]Q[265]Q[129]Q[119]},
  cells = {c},
  hlines,
}
\textbf{ROC AUC} & \textbf{AUC-PR} & \textbf{Brier Score} & \textbf{$R^2\text{m}$}   & \textbf{$R^2\text{c}$}  \\
0.851   & 0.496  & 0.074       & 0.306 & 0.367 
\end{tblr}
\end{table}

\textbf{Our model demonstrates strong discriminatory power, a balanced precision–recall trade-off, and well-calibrated predicted risk estimates.} As shown in Table~\ref{tab:model_performance}, our mixed-effects model achieves an excellent ROC AUC of 0.85, far exceeding the 0.5 expected from random guessing, indicating a strong ability to distinguish between rerun and non-rerun successful jobs. Its AUC-PR of 0.496 substantially exceeds the baseline value of 0.1107, demonstrating effective identification of positive instances while limiting false positives, particularly important in our imbalanced dataset context. Additionally, the low Brier score of 0.074 reflects well-calibrated predictions. Moreover, the marginal $R^2$ of 0.306 (above 0.2) indicates an excellent model fit. This value approximately represents 83\% of the conditional $R^2$ of 0.367, meaning that fixed effects alone account for 83\% of the explained variance in our model. Furthermore, we obtained an EPV of 369, a value far exceeding the threshold of 10, which provides strong evidence that the risk of model overfitting is negligible.

\begin{table}
\caption{Results of Our Mixed-Effects Logistic Model}
\label{tab:feature_importances}
\centering
\begin{scriptsize}
\begin{tblr}{
  width = \linewidth,
  rowsep = .5pt,
  leftsep=2pt,
  colspec = {Q[420]Q[128]Q[108]Q[190]Q[48]Q[60]},
  column{2} = {r},
  column{3} = {r},
  column{4} = {r},
  hline{1-2,19,32,35} = {-}{},
}
\textbf{Factor}                      & \textbf{Coef.} & \textbf{$\chi^2$} & \textbf{Pr($< \chi^2$)} & \textbf{S.$^+$} & \textbf{OR} \\
(Intercept)                          & $-2.806$       & $234.71$    & $< 2.2e^{-16}$ & ***        & $0.06$      \\
job\_type\_code\_and\_image\_build   & $0.124$        & $65.9$      & $4.763e^{-16}$ & ***        & $1.13$      \\
job\_type\_image\_security\_scan     & $-0.281$       & $229.1$     & $< 2.2e^{-16}$ & ***        & $0.75$      \\
job\_type\_notification              & $0.091$        & $43.9$      & $3.403e^{-11}$ & ***        & $1.10$      \\
job\_type\_other                     & $0.031$        & $5.8$       & $0.0163$       & **         & $1.03$      \\
job\_type\_static\_code\_analysis    & $0.399$        & $500.0$     & $< 2.2e^{-16}$ & ***        & $1.49$      \\
job\_type\_testing                   & $0.636$        & $1394.7$    & $< 2.2e^{-16}$ & ***        & $1.89$      \\
branch\_type\_feature                & $-0.642$       & $2063.2$    & $< 2.2e^{-16}$ & ***        & $0.53$      \\
branch\_type\_hotfix                 & $-0.343$       & $730.9$     & $< 2.2e^{-16}$ & ***        & $0.71$      \\
branch\_type\_main                   & $-0.011$       & $1.1$       & $0.3019$       &            & $0.99$      \\
branch\_type\_release                & $-0.243$       & $387.5$     & $< 2.2e^{-16}$ & ***        & $0.78$      \\
month\_of\_year\_sin                 & $0.013$        & $1.2$       & $0.27296$      &            & $1.01$      \\
month\_of\_year\_cos                 & $0.082$        & $42.1$      & $8.627e^{-11}$ & ***        & $1.09$      \\
day\_of\_week\_sin                   & $-0.026$       & $4.8$       & $0.0277$       & *          & $0.97$      \\
day\_of\_week\_cos                   & $-0.010$       & $0.6$       & $0.4218$       &            & $0.99$      \\
hour\_of\_day\_sin                   & $0.003$        & $0.1$       & $0.7917$       &            & $1.00$      \\
hour\_of\_day\_cos                   & $-0.007$       & $0.3$       & $0.5781$       &            & $0.99$      \\%[0.8em]
lang\_gherkin                        & $-0.036$       & $0.2$       & $0.6563$       &            & $0.96$      \\
lang\_go                             & $0.004$        & $0.0$       & $0.9055$       &            & $1.00$      \\
lang\_hcl                            & $0.048$        & $5.8$       & $0.0160$       & *          & $1.05$      \\
lang\_html                           & $-0.153$       & $4.1$       & $0.0429$       & *          & $0.86$      \\
lang\_javascript                     & $0.054$        & $1.4$       & $0.2421$       &            & $1.06$      \\
lang\_jinja                          & $-0.077$       & $0.6$       & $0.4451$       &            & $0.93$      \\
lang\_shell                          & $0.156$        & $11.5$      & $0.0007$       & ***        & $1.17$      \\
lang\_tsx                            & $0.073$        & $0.5$       & $0.4916$       &            & $1.08$      \\
lang\_typescript                     & $-0.177$       & $5.1$       & $0.0236$       & *          & $0.84$      \\
rerun\_ratio                         & $0.273$        & $123.5$     & $< 2.2e^{-16}$ & ***        & $1.31$      \\
failure\_ratio                       & $-0.069$       & $7.0$       & $0.0082$       & **         & $0.93$      \\
build\_frequency                     & $0.001$        & $0.0$       & $0.9644$       &            & $1.00$      \\
commits                              & $0.120$        & $28.3$      & $1.026e^{-07}$ & ***        & $1.13$      \\%[0.8em]
author\_exp\_days                    & $0.008$        & $0.3$       & $0.5891$       &            & $1.01$      \\
author\_exp\_commits                 & $-0.151$       & $85.3$      & $< 2.2e^{-16}$ & ***        & $0.86$      \\
author\_recent\_rerun\_history       & $0.774$        & $6239.2$    & $< 2.2e^{-16}$ & ***        & $2.17$      \\
\SetCell[c=6]{l} {$^+$Significance codes:  0 ‘***’ 0.001 ‘**’ 0.01 ‘*’ 0.05 ‘.’ 0.1 ‘ ’ 1} 
\end{tblr}
\end{scriptsize}
\end{table}

\textbf{The most important factors influencing the likelihood of rerunning successful jobs and associated with higher odds, include job type (especially testing and static code analysis jobs) and projects written in languages with limited IDE assistance such as Shell.} As shown in Table~\ref{tab:feature_importances}, common wisdom factors (i.e., author recent rerun history, rerun ratio, and \#commits) present highly significant evidence (p-value $< 0.001$) of association with the rerun of successful jobs. In particular, the \texttt{\small author recent rerun history} has the highest $\chi^2$ value among all the studied independent variables. 
Its odd ratio suggests that for every one-unit increase, there is a predicted 117\% average increase in the odds of rerunning successful jobs, controlling for all other independent variables.

More interestingly, we observe that several other less obvious factors have a significant influence on the likelihood of rerunning successful jobs. In fact, nearly all dummy variables for job type and branch type show strong evidence of association with success reruns. Specifically, testing and static code analysis jobs, along with feature and hotfix branches, have the four highest $\chi^2$ values among the dummy variables, indicating very high significance.
Based on the odds ratios, successful testing, static code analysis, and code and image build jobs are predicted to have, on average, 89\%, 49\%, and 13\% higher odds of being rerun compared to deployment and release jobs (the reference category), respectively, controlling for all other independent variables. Conversely, successful jobs executed on feature and hotfix branches are 47\% and 29\%, respectively, less likely to be rerun compared to jobs on develop branches.

\begin{figure}
  \begin{center}
      \includegraphics[width=.75\linewidth]{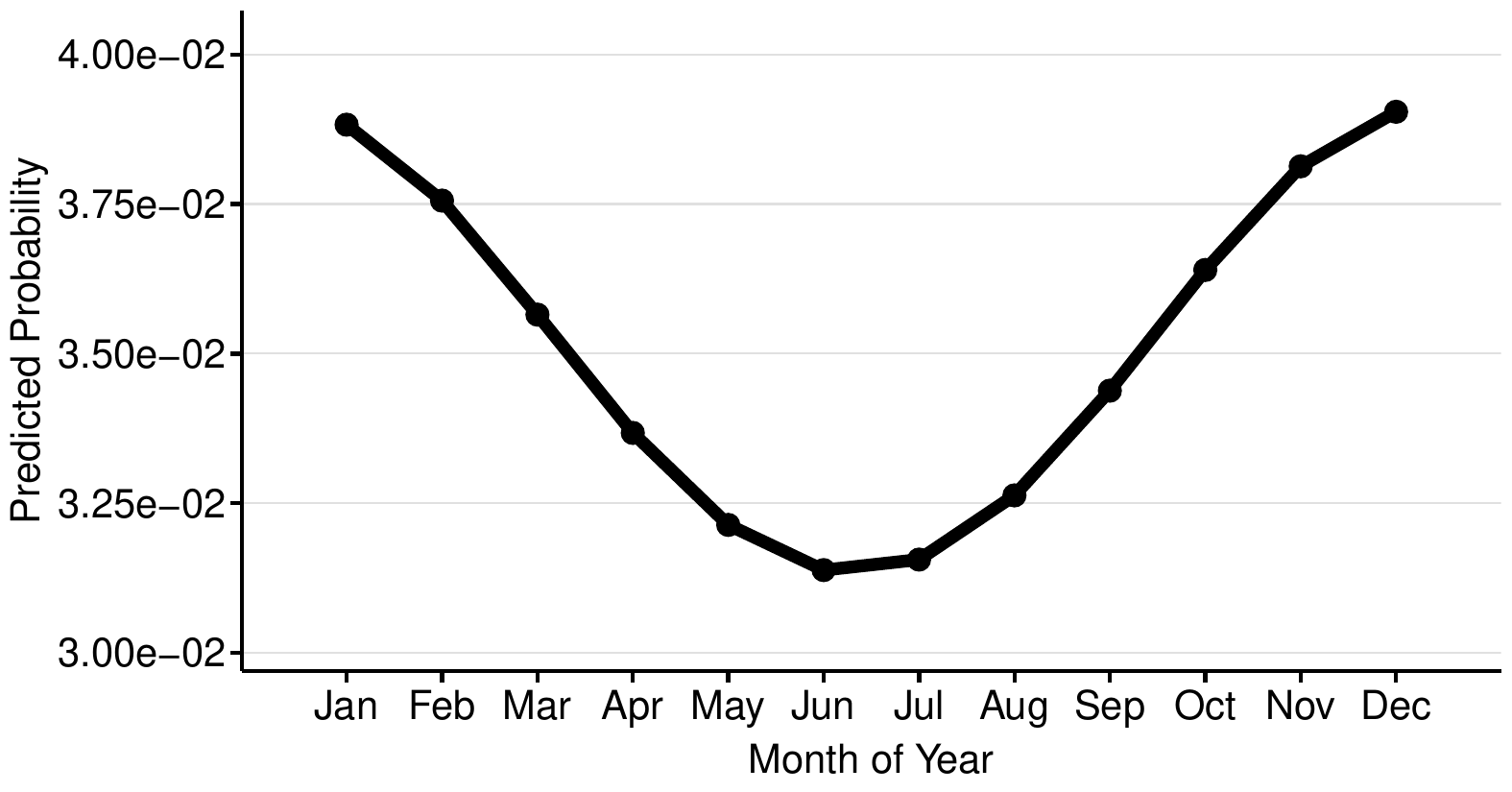}
  \end{center}
\caption{Month of year vs. the probability of rerunning successful jobs}
\label{fig:month_prob}
\end{figure}

Among the periodic variables, the cosine component of the month of year is highly significant (p-value $< 0.001$), suggesting a seasonal trend. To interpret its influence, we plot in Fig.~\ref{fig:month_prob} its response curve to analyze how the predicted probability of rerunning successful jobs evolves as the months vary, controlling for all other variables. As can be observed, successful jobs executed at the end or beginning of the year are predicted to have higher odds of being rerun, on average, compared to those executed in the middle of the year. \added{While this seasonal trend is statistically significant ($\chi^2 = 42.1$, p-value $< 0.001$), we note that its practical effect size is modest (OR = 1.09) compared to factors such as job type (OR up to 1.89) or author recent rerun history (OR = 2.17). We retain temporal features because they reveal a meaningful pattern for monitoring purposes, but practitioners should prioritize addressing job-type and author-level factors for greater impact.}

At the project level, we note that both Shell and HCL languages are significant and associated with higher odds. In particular, successful jobs executed in Shell and HCL (Infrastructure-as-Code) projects are predicted to have, on average, respectively 17\% and 5\% higher odds of being rerun than those in Python projects, controlling for all other independent variables. \added{The higher odds for Shell and IaC-heavy projects could reflect multiple factors, including their limited IDE support for catching bugs, their prevalence in tasks that interface with external systems (caches, artifact repositories, deployment environments), and the common scripting practice of not enforcing strict exit code checking (e.g., omitting \texttt{set -e}) that allow failures to go undetected. 
These explanations remain hypothetical and warrant further investigation to establish causality.}

\textbf{Other key factors associated with lower odds of rerunning successful jobs, include execution context (e.g., feature, hotfix, and release branches), project failure ratio, and developer experience in commits contribution.} As shown in Table~\ref{tab:feature_importances}, successful jobs executed on feature, hotfix, and release branches are highly significant (p-value $< 0.001$) and predicted to have, on average, respectively 47\%, 29\% and 22\% lower odds of being rerun compared to those executed on develop branches. 
\added{These results are consistent with the GitFlow branching model adopted at TELUS (see Section~\ref{sec:background}). As the central integration branch, \texttt{develop} is where all feature branches converge, leading to a higher frequency of CI jobs and a greater exposure to unexpected issues such as caching conflicts, artifact inconsistencies from parallel contributions, and environmental instabilities. In contrast, \texttt{feature} branches represent more isolated development contexts, where a single developer typically works independently on a specific change. This isolation reduces exposure to the types of intermittent and integration-related problems that can cause silent failures, which may explain the 47\% lower odds observed for feature branches. Similarly, \texttt{hotfix} and \texttt{release} branches focus on well-defined, targeted changes (urgent fixes and stabilization, respectively), further reducing the likelihood of encountering unexpected silent issues compared to the more dynamic \texttt{develop} branch.}
Similarly, for every one-unit increase in the project's failure ratio (p-value $< 0.05$) there is a predicted 7\% average lower odds of successful jobs being rerun. \removed{This suggests that projects with higher failure ratios have more effective feedback mechanisms, resulting in fewer unnoticed failures that would prompt the rerun of successful jobs.}\added{This association may indicate that projects with higher failure ratios benefit from more explicit error signals, potentially reducing the occurrence of unnoticed failures that would otherwise prompt the rerun of successful jobs. However, this interpretation remains a hypothesis that warrants further investigation.} Finally, successful jobs authored by developers with greater commit contributions to the project, tend to have lower odds of being rerun. Specifically, for each one-unit increase in the commit ratio of a successful job's author, the odds of its rerun are predicted on average to decrease by 14\%, holding all other independent variables constant.

\added{\textbf{Replication of RQ\textsubscript{2} on OSS projects.} Our replication on 8 OSS projects (Online Appendix~F\cref{note:online_appendix}) yielded a model with even stronger discriminatory power (ROC AUC of 0.976, AUC-PR of 0.849, Brier Score of 0.021), confirming that the rerunning of successful jobs is driven by distinct factors. Several variables exhibit consistent trends across both contexts, including testing and static code analysis job types being associated with higher rerun odds. However, some variables show opposite effects (e.g., \texttt{commits}), which we attribute to contextual differences between industrial microservices and larger OSS projects. These results support the generalizability of our core finding that specific, identifiable factors are associated with the rerun of successful jobs.}

\begin{tcolorbox}[rqbox={Summary of RQ\textsubscript{2}}]
\removed{Successful jobs are more likely to be rerun when they are testing and static code analysis jobs, or belong to projects in languages with limited IDE support, such as Shell and HCL.}\added{Successful jobs are more likely to be rerun when they are code or image build, testing, and static code analysis jobs, or belong to projects dominated by scripting and infrastructure-as-code languages such as Shell and HCL.} In contrast, the odds of rerunning successful jobs decrease when they are on feature, hotfix, or release branches, in projects with higher failure ratios, or triggered by more experienced contributors.
\end{tcolorbox}

 \subsection{\textbf{\rqthree}}
\label{sec:rq3}

\textbf{Motivation.} In this third RQ, we seek to explore the root cause of silent job failures and existing solutions discussed by developers. The findings will provide valuable insights, complementary to the key influential factors identified in RQ\textsubscript{2}, to help practitioners better understand the nature of silent failures that occur during job executions marked as successful and find initial solutions to these issues.

\textbf{Approach.} We conduct a thematic analysis of practitioner-reported issues on GitLab. We focus on publicly reported issues to gain broader insights beyond the studied industrial context and to ensure the generalizability of our findings. Inspired by prior work \cite{weeraddana_characterizing_2024}, we first systematically search for relevant issues related to silent failure.

\added{The search queries used include:
\begin{quote}
    \begin{itemize}[nosep]
      \item {silent (job \texttt{OR} build \texttt{OR} pipeline \texttt{OR} CI) failure}
      \item {(job \texttt{OR} build) succeeds but fails}
      \item {(job \texttt{OR} build) fails but still succeeds}
    \end{itemize}
\end{quote}
}
\removed{``\texttt{silent (job OR build OR pipeline OR CI) failure}", ``\texttt{(job OR build) succeeds but fails}", ``\texttt{(job OR build) fails but still succeeds}" use on the GitLab issue tracker} 
\added{These queries capture our definition of silent job failures corresponding to jobs that are marked as successful but in reality fail to complete their task.}
\added{Our search was conducted on the official GitLab issue tracker\footnote{\url{https://gitlab.com/gitlab-org/gitlab/-/issues}}, which covers both the GitLab platform and the GitLab Runner projects. We focused on this single data source because (1)~our study is centered on GitLab CI, the most widely adopted self-hosted CI platform, (2)~practitioners commonly report CI-related issues on the official tracker, and (3)~restricting to one platform ensures consistency in the taxonomy.}
We intentionally run our search on all issues regardless of status (i.e., open/closed) or recency. Indeed, practitioners are often reluctant to make major migrations \cite{aidasso_diagnosis_2025} and still rely on older versions of tools, and therefore, may still encounter or learn from previously resolved problems. Hence, we manually screen issues in the search results, examining their titles and descriptions to retain only those explicitly describing scenarios where failures occurred during CI job execution, but the job was ultimately marked as successful.

This filtering process resulted in 92 relevant issues identified, including 1,199 human-written comments. Then, using open coding \cite{khandkar_open_2009}, we examine in a first iteration, the title, description, and comment threads of each issue to identify recurring themes (or codes) related to the causes and contexts of the reported silent failures. At the end, we further refine the identified themes by grouping similar topics and renaming themes to account for their new scope. In a second iteration, the first two authors \added{(i.e., two coders)} independently assigned themes to issues, resolving disagreements through discussion. To assess the consistency of our categorization, we calculated Cohen’s Kappa score \cite{cohen_weighted_1968}, which yielded a value of 0.913, indicating almost perfect agreement. In addition, we identified and summarized the solutions or workarounds proposed by practitioners and CI provider engineers for each theme. Hence, we provide insights into both the problem space and available remediation strategies. The detailed list of the 92 issues with their links is included in our replication package \cite{aidasso_replication_2025}.

\stepcounter{footnote}
\footnotetext{\label{note:security_scan_failure_ex2}\url{https://gitlab.com/gitlab-org/gitlab/-/issues/347479}}

\stepcounter{footnote}
\footnotetext{\label{note:post_deploy_ex1}\url{https://gitlab.com/gitlab-org/gitlab/-/issues/16586}}

\stepcounter{footnote}
\footnotetext{\label{note:ignored_test_ex1}\url{https://gitlab.com/gitlab-org/gitlab/-/issues/501738}}

\stepcounter{footnote}
\footnotetext{\label{note:silent_curl_error_ex}\url{https://gitlab.com/gitlab-org/charts/gitlab/-/issues/3145}} 

\stepcounter{footnote}
\footnotetext{\label{note:invalid_ci_variables_ex}\url{https://gitlab.com/gitlab-org/gitlab/-/issues/556631}}

\stepcounter{footnote}
\footnotetext{\label{note:syntax_error_ex}\url{https://gitlab.com/gitlab-org/gitlab/-/issues/340541}}

\begin{table*}
\centering
\caption{Themes and Solutions Identified for Silent Job Failures}
\label{tab:themes}
\begin{scriptsize}
\begin{tblr}{
  width = \linewidth,
  rowsep = 3pt,
  leftsep=2pt,
  colspec = {Q[30,m]Q[190]Q[400]Q[600]},
  hlines,
  cells = {m},
}
\textbf{ID} & \textbf{Theme (\#)}  & \textbf{\added{Description \& Example}}   & \textbf{Solutions or workarounds}                                                                                                                                                                                                                                                           \\
T{1}          & Artifact Operation\hfill\break Errors (26) & \added{Failures in artifact upload, download, or retrieval that are silently ignored. E.g., a project named \texttt{job\_should\_fail}\cref{note:artifact_issue_1} demonstrates a job that succeeds despite failing to upload declared artifacts, while another\cref{note:artifact_issue_2} shows a JUnit report upload failing silently with no error trace.}
     & {- Add dependency between artifact-producing and consuming jobs.\\- Increase the maximum artifact size for uploads in settings\\- Fix local paths and permanent storage configurations\\- Check artifacts' presence and explicitly fail job} \\
T{2}          & Caching\hfill\break Errors (21)     & \added{Cache upload/download failures (connection errors, empty keys, expired files) that produce no warning. E.g., the log excerpt in Fig.~\ref{fig:silent_failure_log1} shows a cache upload that fails silently in a successful job\cref{note:cache_error_example_1}, with the \texttt{Created cache} message appearing despite the actual upload failing.}
           & {- Check storage access and retention policy configurations\\- Clean up cache storage disk space\\- Commit cache files separately to avoid overrides on same commit ID\\- Upgrade the cache management system version}                                            \\
T{3}          & Ignored Non-Zero\hfill\break Exit Codes (17) & \added{Job scripts return non-zero exit codes or execute \texttt{exit 1}, but the job still reports success. E.g., an \texttt{exit 1} inside an \texttt{if-else} block of a PowerShell script is silently swallowed\cref{note:explicit_exit_1}, or a failed \texttt{rsync} deployment (\texttt{exit code 255}) is ignored\cref{note:rsync_deploy_failed}.}
  & {- Set \texttt{GIT\_STRATEGY=clone} in CI config. to fix failed repository cloning issues\\- Avoid \texttt{exit} calls in if-else blocks of Powershell scripts~\\- Add~ \texttt{set -e} to scripts to enforce exit code checks\\-~Use {source} ./script.sh instead of ./script.sh in Dockerfiles}       \\
T{4}          & Infrastructure\hfill\break Config. Issues (10) & \added{CI infrastructure misconfigurations (including, restrictive worker limits, k8s pod terminations) that silently abort jobs. E.g., k8s-managed runners abruptly terminate jobs mid-execution (\texttt{Error cleaning up pod} in logs) while still reporting success\cref{note:k8s_bug} (\texttt{Job succeeded}).}
 & {- Increase the worker timeout\\- Check or increase worker task queue size limits\\-~Set pod limits and requests in k8s managed runners}                                                                                                                                             \\
T{5}          & Security Scan\hfill\break Failures (7) & \added{Security or license scans fail (e.g., missing analyzer dependencies) but the job signals success. E.g., a license scanning job ``succeeds, but actually fails silently without reporting anything''\cref{note:security_scan_failure_ex1}\textsuperscript{,}\cref{note:security_scan_failure_ex2}.}
    & {- Install missing analyzer deps. before running the script\\- Upgrade or change the scanning tool or library}                                                                                                                                                                       \\
T{6}          & Skipped Script\hfill\break Execution (7) & \added{Job scripts or triggers are partially/completely skipped, yet the job is marked successful. E.g., misconfigured container image entrypoints cause the entire CI configuration to be bypassed\cref{note:skipped_script_exec_ex1}.}
      & {- Check overrides of container image entrypoints\\- Update runner version from this list \cite{noauthor_windows_2021}}                                                                                                                                                                                              \\
T{7}          & Post-deployment\hfill\break Issues (6) & \added{Deployment jobs succeed but the environment is not actually created/updated or the deployed container is unavailable. E.g., deployment environment creation fails silently due to an incorrect environment name or URL\cref{note:post_deploy_ex1}\textsuperscript{,}\cref{note:post_deploy_ex2}.}
         & {- Check deployment env. configs. (e.g., env. name, URL)\\- Add post-deployment steps to check container health}                                                                                                                                 \\
T{8}          & Ignored Test\hfill\break Failures (3) & \added{Test cases fail but results and non-zero exit codes are ignored, allowing the job to pass. E.g., a WinBatch test script fails but the job succeeds because the error level is not properly checked\cref{note:ignored_test_ex1}.}
         & {- Impl. post-testing steps to check failed test cases in reports\\- Explicitly set proper error level in WinBatch test scripts}                                                                                                                                    \\
T{9}          & Silent Curl\hfill\break Errors (3)  & \added{Use of \texttt{curl --silent} (\texttt{-s}) suppresses error output, masking failures in artifact uploads or HTTP requests. E.g., a \texttt{curl -s} command fails to upload an artifact but no error is shown\cref{note:silent_curl_error_ex}.}
            & {- Remove the  –silent or  -s option that silences curl errors\\- Add an  OR command to the problematic script to send email notifications in case of failures}                                                                                                                                          \\
T{10}         & Invalid CI\hfill\break Variables (2) & \added{Misconfigured CI/CD variables (wrong scopes, self-referencing without expansion) cause silent misbehavior. E.g., a variable scoped only to protected branches silently evaluates to empty on other branches\cref{note:invalid_ci_variables_ex}.}
       & {- Check variables scopes (e.g., protected branches only)\\- Enable/disable variable expansion}                                                                                                                                                                                             \\
T{11}         & Syntax Errors (1) & \added{Partial CI configurations saved in separate files bypass YAML validation, allowing syntax errors to cause silent failures\cref{note:syntax_error_ex}.}
     & - Avoid including partial CI configurations in separate files                                                                                                                                      
\end{tblr}
\end{scriptsize}

\end{table*}

\begin{table}[h]
\caption{Cross-tabulation of silent failure themes and job types$^{\dagger}$ identified from issue descriptions.}
\label{tab:theme_jobtype}
\centering
\begin{scriptsize}
\begin{tblr}{
  width = \linewidth,
  colspec = {Q[230] *{11}{Q[45]} Q[90]},
  cells = {c},
  column{1} = {l},
  hline{1-2,9,10} = {-}{},
  row{9} = {font=\bfseries},
}
\textbf{Job type} & \textbf{T1} & \textbf{T2} & \textbf{T3} & \textbf{T4} & \textbf{T5} & \textbf{T6} & \textbf{T7} & \textbf{T8} & \textbf{T9} & \textbf{T10} & \textbf{T11} & \textbf{Total} \\
Build & 15 & 9 & 6 & 2 & - & 3 & - & - & 1 & - & - & 36 \\
Test & 7 & 8 & 8 & 1 & - & 1 & - & 3 & - & - & - & 28 \\
Deploy & - & 1 & 1 & 1 & - & 1 & 6 & - & - & 1 & - & 11 \\
Security & 1 & - & - & - & 7 & - & - & - & - & - & - & 8 \\
SCA & 3 & 2 & - & 1 & - & - & - & - & - & - & - & 6 \\
Notif & - & - & - & - & - & - & - & - & 1 & - & - & 1 \\
Unspec & 8 & 8 & 3 & 6 & - & 2 & - & - & 1 & 1 & 1 & 30 \\
Total & 26 & 21 & 17 & 10 & 7 & 7 & 6 & 3 & 3 & 2 & 1 & 103 \\
\end{tblr}
\\[2pt]
\raggedright\footnotesize $^{\dagger}$Build: \texttt{code\_and\_image\_build}, Test: \texttt{testing}, Deploy: \texttt{deployment\_and\_release}, Security: \texttt{image\_security\_scan}, SCA: \texttt{static\_code\_analysis}, Notif: \texttt{notification}, Unspec: \texttt{unspecified}. 
\end{scriptsize}
\end{table}

\added{In addition, we coded each issue for the job type(s) associated, using the same seven types defined in Section~\ref{sec:explanatory_factors}. Job types were identified from explicit mentions in issue descriptions and comments, or inferred from referenced artifacts (e.g., coverage reports $\rightarrow$ testing, DAST reports $\rightarrow$ security scanning). Issues where no specific job type could be determined were marked as ``unspecified.'' An issue could be assigned multiple job types when applicable.}

\textbf{Results.} Table~\ref{tab:themes} summarizes the 11 identified themes along \added{with brief descriptions, concrete examples drawn from the referenced public issues}, and proposed solutions or workarounds. We elaborate on the themes below.

\textbf{T1. Artifact Operation Errors.} This theme, the most commonly observed, includes failures related to artifact handling, such as uploading, downloading, and unexpected issues like retrieving outdated artifact versions. As noted in issues,\footnote{\label{note:artifact_issue_1}\url{https://gitlab.com/gitlab-org/gitlab/-/issues/22711}} \footnote{\label{note:artifact_issue_2}\url{https://gitlab.com/gitlab-org/gitlab/-/issues/33566}} developers expect these errors to cause immediate job failure rather than being suppressed, since they can lead to serious problems, ranging from downstream job failures to bugs reaching production.\cref{note:artifact_issue_1} Artifacts often include critical resources such as unit test or security reports, making it essential that upload or retrieval failures are noticed right away.\cref{note:artifact_issue_2}

\textbf{T2. Caching Errors.} 
The second most common theme involves cache-related failures, such as connection errors, ignored empty keys, and issues with expired or incomplete cache files, that occur without feedback to developers. These problems often stem from local cache misconfigurations,\footnote{\url{https://gitlab.com/gitlab-org/gitlab/-/issues/291052}} connection issues with external storage like Google Cloud Storage,\footnote{\url{https://gitlab.com/gitlab-org/gitlab-runner/-/issues/4127}} and concurrency conflicts in parallel jobs sharing the same cache.\footnote{\label{note:cache_waste}\url{https://gitlab.com/gitlab-org/gitlab/-/issues/291052}} These silent failures cause downstream job failures due to missing or corrupted caches, forcing manual reruns of early cache-producing jobs, a process that involves up to 4 wasted minutes on each cache creation step\cref{note:cache_waste}.

\textbf{T{3}. Ignored Non-Zero Exit Codes.} Another common theme includes cases where, either a job script returns an error code\footnote{\label{note:rsync_deploy_failed}\url{https://gitlab.com/gitlab-org/gitlab-runner/-/issues/26877}} (i.e., a non-zero code) or an explicit failure command is executed\footnote{\label{note:explicit_exit_1}\url{https://gitlab.com/gitlab-org/gitlab-runner/-/issues/26538}} (e.g., \texttt{\small exit code 1}), but the job ignores this non-zero exit code and still reports success to the developers. Such an issue has been reported as ``driving [\dots] team crazy after weeks searching for a fix''\cref{note:explicit_exit_1} and can lead to highly problematic scenarios where deployments are unsuccessful but remain unnoticed\cref{note:rsync_deploy_failed} for long periods. Migrating to a non-buggy runner version has also been suggested \cite{noauthor_powershell_2023}.

\removed{\textit{Removed detailed description of T4 to T11}}

\added{The remaining themes (T4 to T11) are summarized in Table~\ref{tab:themes} with descriptions, concrete examples, and proposed solutions. We highlight several aspects that underscore the practical severity of silent failures. Infrastructure configuration issues (T4), such as Kubernetes clusters abruptly terminating jobs mid-execution,\footnote{\label{note:k8s_bug}\url{https://gitlab.com/gitlab-org/gitlab-runner/-/issues/4119}} can force developers to ``manually restart the pipelines several times an hour''\cref{note:k8s_bug}, resulting in significant waste. Skipped script execution (T6) has been described as a problem ``which can [have] quite disastrous results [in production]'',\footnote{\label{note:skipped_script_exec_ex1}\url{https://gitlab.com/gitlab-org/gitlab/-/issues/13000}} as teams may falsely certify code quality or announce releases without any actual deployment. Post-deployment issues (T7) include cases where containers are marked as successfully deployed but are not actually available in the target environment.\footnote{\label{note:post_deploy_ex2}\url{https://gitlab.com/gitlab-org/gitlab-runner/-/issues/25988}} Ignored test failures (T8) pose a critical risk, as failing tests that go undetected can lead to buggy code reaching production. The remaining themes, i.e., security scan failures (T5), silent curl errors (T9), invalid CI variables (T10), and syntax errors (T11), are less frequent but further illustrate how misconfigurations and tool misuse can mask failures under a success signal.}

\added{To explore the connection between the identified silent failure themes and the influential factors of successful job reruns from RQ\textsubscript{2}, we cross-tabulate themes with the job types reported in each issue. We focus on job type because it is the most significant categorical factor in RQ\textsubscript{2} and the most reliably identifiable from issue reports, unlike factors such as branch type or developer experience. Job type was identifiable in 64 of 92 issues (70\%). Table~\ref{tab:theme_jobtype} presents the results.}

\added{Several patterns emerge that converge with the RQ\textsubscript{2} findings. Build and testing jobs concentrate in the three most frequent themes: T1. Artifact Operation Errors (22 of 26), T2. Caching Errors (17 of 21), and T3. Ignored Non-zero Exit Codes (14 of 17). This is consistent with the RQ\textsubscript{2} finding, where testing and static code analysis jobs exhibit the highest rerun odds (OR of 1.89 and 1.49, respectively), as these job types are the primary producers and consumers of artifacts and caches. As expected, Security scan failures (T5) map exclusively to security scan jobs (7 of 7), Post-deployment issues (T7) exclusively to deployment and release jobs (6 of 6), and Ignored test failures (T8) exclusively to testing jobs (3 of 3). Nevertheless, Infrastructure config. issues (T4) are predominantly unspecified (6 of 10), reflecting their cross-cutting nature as platform-level problems. Overall, the concentration of build and testing jobs in the most frequent silent failure themes aligns with their higher rerun odds in RQ\textsubscript{2}, providing empirical support for the connection between the quantitative and qualitative findings.}

\begin{tcolorbox}[rqbox={Summary of RQ\textsubscript{3}}]
We identified 11 themes related to silent failures, including cache and artifact operation issues (T{1}, T{2}); failures within \removed{build script}\added{CI configuration file}s, such as ignored test and security scans (T{3}, T{5}, T{8}); infrastructure problems that disrupt job completion (T{4}, T{6}); and misconfigurations or misuse of CI tools (T{7}, T{9}, T{10}, T{11}) leading to missed alerts, escaped bugs, and unnoticed deployment failures.
\end{tcolorbox}

\section{Implications}
\label{sec:implications}

\textbf{Practitioners should include post-execution verification tasks as a standard step in CI pipelines to validate the success of business-critical jobs that produce artifacts or perform environment changes such as deployments.}
As determined in RQ\textsubscript{2}, the most important factors influencing the rerun of successful jobs include specific job types that either produce artifacts (e.g., testing and static code analysis jobs) or perform environmental changes (e.g., notification and image build jobs). These findings align perfectly with the most common causes of silent failures identified in RQ\textsubscript{3}, including artifact operation errors and caching errors, but also ignored test failures and post-deployment issues, which do not resulting in actual job failures, even though developers expect such issues to be explicitly surfaced. Indeed, these silent failures can have catastrophic consequences for organizations. As an example, falsely successful testing jobs can lead to bugs escaping into production, affecting software quality, user experience, and ultimately the organization's reputation. To anticipate these problems, we recommend that practitioners systematically add verification steps (e.g., jobs) immediately after jobs whose successful and reliable executions are critical to the business, in order to ensure the presence of the expected artifacts or environmental changes (e.g., test reports, cache updates, new container image or software versions). In this way, these downstream jobs can detect silent failures and provide timely feedback to developers through explicit failures.

For the specific studied industrial context, we observed additional factors that can help raise awareness of silent failures across teams to anticipate and mitigate them. In particular, 
\removed{job specifications in projects mainly written in Shell or HCL warrant attention to avoid common pitfalls (e.g, exit code 1 in an \textit{if-else} block) so as to improve reliability.}
\added{the association between Shell/HCL projects and higher rerun odds may stem from multiple factors, including limited IDE assistance for early error detection and common scripting pitfalls such as missing \texttt{set -e} directives and ignored exit codes in \textit{if-else} blocks (see T3 and T9 in RQ\textsubscript{3}). Teams working with scripting and infrastructure-as-code languages should review their scripts for such patterns to improve reliability.}

Also, we found that less experienced developers rerun successful jobs more often, which might indicate the need for better onboarding and assistance from the more experienced developers regarding unreliable outcomes. Furthermore, the load period from November to February is associated with higher odds of rerunning successful jobs. While further investigation is required to understand the underlying reasons for this tendency (e.g., end-of-year holiday effects or scheduled system updates), such information is a useful starting point for efficient monitoring.

\added{In summary, based on the findings of this study, \textbf{we recommend that practitioners adopt the following measures to mitigate silent failures}:}
\added{\begin{itemize}[nosep]
\item Add post-execution verification steps after artifact-producing and deployment jobs (including build, testing, and static analysis jobs) to validate expected outputs.
\item Enforce \texttt{set -e} in Shell scripts and review exit code handling in conditional blocks.
\item Periodically review cache configurations and artifact retention policies on self-hosted CI instances.
\item Prioritize monitoring of testing and static code analysis jobs, which exhibit the highest rerun odds.
\item Support less experienced developers through onboarding on CI reliability practices.
\end{itemize}}

\textbf{Our findings highlight important root causes that CI providers should prioritize to prevent silent failures and enhance the reliability of CI systems.}
In this study, we identified several infrastructure-related factors that influence the rerun of successful jobs, as well as the most common causes of silent failures that prompt developers to trigger such reruns. In particular, caching issues, artifact operation errors, and ignored failure exit codes are the most common, but can be mitigated through enhanced validation mechanisms that we recommend CI providers implement in future versions. For example, they could introduce in the \removed{build script}\added{CI configuration file} a notion of “required artifacts”, whose presence would be systematically verified after job execution, with explicit failure raised if they are missing. Similarly, CI providers should enable configuration options that clearly signal failures related to cache or artifact repository storage access, rather than silencing them in seemingly successful jobs. As part of the root causes, we also identified several CI configuration smells (e.g., short worker timeout, small queue size, or partial configuration file saved outside the \removed{build script}\added{CI configuration file}). We recommend that these smells be leveraged to improve validation of \removed{build script}\added{CI configuration file}s, both within IDEs and at submission time, to prevent silent failures.
\added{In this regard, several of our RQ\textsubscript{3} themes, particularly Ignored Non-zero Exit Codes (T3), Invalid CI Variables (T10), and Syntax Errors (T11), can be viewed as CI configuration smells that produce fake successes, as conceptualized by Vassallo et al.~\cite{vassallo_configuration_2020}. Extending linting tools such as CD-Linter~\cite{vassallo_configuration_2020} with detection rules targeting these patterns can help anticipate a subset of the silent failures we identified.}

\textbf{Researchers should propose approaches for automated detection of silent failures, to assist developers in identifying them early, so as to reduce the costs of job reruns.}
Our mixed-effects model of jobs from 81 projects shows strong discriminatory power (ROC AUC of 85\%), is well calibrated (Brier score of 0.074) and achieves an excellent goodness-of-fit score (marginal $R^2$ of 0.306). \removed{It thus demonstrates that the studied variables can be useful in predicting whether a successful execution is likely to be rerun (i.e., due to a silent failure) or not. However, our main objective in this study was to investigate the most influential factors associated with these reruns rather than to predict future reruns of successful jobs.}
\added{These results suggest that the studied variables capture meaningful variation between rerun and non-rerun successful jobs. We emphasize that the primary goal of our model is \emph{explanatory}: i.e., to identify the most influential factors associated with the rerun of successful jobs. We did not design, tune, or evaluate the model for predictive deployment (e.g., no train/test split or temporal validation was used).} 
\removed{Nonetheless, we invite future studies to leverage our findings, particularly the most important factors and common root causes, to develop models for detecting silent failures.}
\added{Nonetheless, our explanatory findings could serve as a starting point for future predictive work. We invite future studies to leverage the identified factors and root causes to develop dedicated detection models for silent failures, using appropriate predictive evaluation protocols (e.g., temporal cross-validation).}

Similar to previous work on models for detecting intermittent failures \cite{lampel_when_2021, olewicki_towards_2022, aidasso_efficient_2025}, these models can help developers identify silent issues early so they can mitigate them before changing context. We also expect these models to reduce the number of reruns of successful jobs, thereby cutting down on the costs associated with the job reruns in CI.

\section{Related Work}
\label{sec:related}

\textbf{Build Failure Factors}. Understanding the factors associated with pipeline outcomes has been a major focus of CI research \cite{aidasso_build_2025}, with most studies conducted on OSS projects. Early work \cite{seo_programmers_2014, maes-bermejo_revisiting_2022} used the open coding \cite{khandkar_open_2009} technique to determine the root causes of build failures, finding that compiler errors (e.g., dependency issues, syntax errors, and missing repository files) and test failures were the most prevalent. Other studies \cite{islam_insights_2017, rausch_empirical_2017} reached similar conclusions using statistical tests, such as the $\chi^2$ test of independence. In particular, Islam et al. \cite{islam_insights_2017} found that job complexity, build strategy (e.g., branch type), and project characteristics like project and team sizes significantly influence build outcomes. Besides change complexity and test failures, Rausch et al. \cite{rausch_empirical_2017} showed that build failures are also strongly associated with build history (i.e., high recent failure rate) and the author’s experience in OSS Java projects. Several further studies~\cite{dimitropoulos_continuous_2017, wrobel_using_2023, luo_what_2017, barrak_why_2021} leveraged ML techniques to explore factors associated with build outcomes, consistently highlighting code complexity, build history, and author experience as important factors. Benjamin et al. \cite{benjamin_study_2023} have recently suggested that contextual project factors alone largely explain build failures and could be used to raise context awareness and focus on projects that are more prone to build failures. Understanding build failure factors has been foundational to multiple studies conducted to address build failures \cite{aidasso_build_2025}. However, the proposed solutions see limited adoption in practice \cite{aidasso_build_2025}, which is likely due to divergent pressing issues in industrial CI environments. In fact, Vassallo et al. \cite{vassallo_tale_2017} showed that OSS and industrial CI contexts differ in that OSS build failures are mainly due to test failures, whereas in industry, deployment and release issues also constitute a major challenge.

\textbf{Build Performance Factors}. Several other studies \cite{silva_what_2023, ghaleb_empirical_2019, weeraddana_characterizing_2024} investigated factors influencing different pipeline performance aspects to improve CI. In particular, Silva et al. \cite{silva_what_2023} used association rules to analyze how factors related to developer activity, project characteristics, and build complexity affect build failure correction time in 18 industrial projects. They found that experienced developers fix failures faster, while complex changes take longer to resolve. Ghaleb et al. \cite{ghaleb_empirical_2019} studied the factors influencing the build duration using mixed-effects regression on 104,442 builds from 67 GitHub projects, and identified internal reruns of failed commands, caching, and execution periods as the most influential ones. More recently, Weeraddana et al. \cite{weeraddana_characterizing_2024} sought to characterize timeout builds in 24 OSS projects, with factors related to build history, job characteristics, timeout tendency, and author experience, using statistical logistic regression models. As a result, they found the build history and timeout tendency as the most influential factors of build timeouts.

\added{Overall, these studies established a common set of feature families, including author experience, project characteristics, job type, and build history, for modeling various CI outcomes (e.g., result, duration, timeout). Our study builds on these foundations by applying similar feature families to a previously unexplored dependent variable: the rerun of \emph{successful} jobs as a proxy for silent failures.}

\added{\textbf{Noise in Build Outcomes.}}
\added{Gallaba et al.~\cite{gallaba_noise_2018} studied noise and heterogeneity in historical build data on Travis CI, revealing that passing builds frequently contain failing or skipped jobs that are actively or passively ignored by developers. Their findings demonstrate that build-level outcomes do not always reflect the actual status of individual jobs, challenging the common assumption of clean pass/fail signals used to train predictive models. While their work operates at the build level (passing builds masking failed jobs), our study focuses on the job level, i.e., individual jobs marked as successful that fail to complete their intended tasks. Together, these findings highlight a broader phenomenon of misleading success signals across multiple granularities of CI.}

\added{\textbf{CI Configuration Smells.} Vassallo et al.~\cite{vassallo_configuration_2020} proposed CD-Linter, a tool for automatically detecting configuration smells in GitLab CI pipelines. Among the four smells identified, the ``fake success'' smell (where configurations such as \texttt{allow\_failure: true} mask genuine errors) is closely related to our notion of silent failures. However, CD-Linter focuses on statically detectable anti-patterns in CI configuration files, whereas our study empirically characterizes the broader phenomenon of silent failures through the lens of job reruns, identifying influential factors and 11 thematic categories that extend beyond configuration smells to include infrastructure issues, scripting pitfalls, and tool-specific bugs.}

\textbf{Intermittent Job Failures.} A notable line of work~\cite{lampel_when_2021, olewicki_towards_2022, aidasso_diagnosis_2025} on CI reliability has focused on intermittent job failures, which are typically due to infrastructure issues or flaky tests~\cite{aidasso_diagnosis_2025, olewicki_towards_2022}, do not provide relevant code-related feedback to developers, but rather mislead them. To support developers in identifying these failures, several studies~\cite{lampel_when_2021, moriconi_automated_2022, olewicki_towards_2022} investigated ML techniques to automate their detection, obtaining promising results at Mozilla~\cite{lampel_when_2021}, Amadeus~\cite{moriconi_automated_2022}, and Ubisoft~\cite{olewicki_towards_2022}. Aïdasso et al.~\cite{aidasso_efficient_2025, aidasso_predicting_2026} leveraged few-shot fine-tuned small language models to further improve detection and diagnosis at TELUS. While these studies address an important CI reliability challenge, they focus solely on explicitly failed jobs. Such failures are arguably easier to identify than silent ones, as they at least notify developers of the presence of a problem.

\added{\textbf{Rerun and Recheck Practices.} Durieux et al.~\cite{durieux_empirical_2020} studied rerun practices in OSS projects on Travis CI and found that rerun-initiating jobs include 17\% and 48\% of successful and failed jobs, respectively. The authors focused on failed reruns to characterize intermittent job failures according to programming languages, failure reasons, and time delays. Maipradit et al.~\cite{maipradit_repeated_2024} conducted an empirical study of 66,932 code reviews from the OpenStack community, finding that 55\% of code reviews invoke the recheck command after a failing build, but only 42\% of rechecks change the build outcome, producing up to 187.4 compute years of waste. Brus et al.~\cite{brus_rechecking_2025} extended this line of work by fitting statistical models to discriminate between successful and failing rechecks across 314,947 recheck requests, showing that past behaviour of jobs, bots, and users are the most important explanatory factors. These studies focus on reruns or rechecks triggered after \emph{explicit CI failures}. Our study addresses the complementary phenomenon: reruns triggered after \emph{apparent success}, where the developer suspects an undetected issue. This distinction maps to two failure modes of the CI signal: false negatives (flaky failures) versus false positives (silent failures). Furthermore, these works operate at the build or change-set level, whereas our analysis is at the job level within GitLab CI pipelines.}

This study focuses on the phenomenon of silent failure, where a job signaling success involves a partial or complete failure to execute its tasks. \removed{To the best of our knowledge, no prior study has investigated CI reliability in this regard.} \added{While existing studies investigate reruns of \emph{failed} jobs, to the best of our knowledge, no prior work has investigated CI jobs marked as successful but failing to complete their intended tasks, i.e., 
silent failures.} To address this gap, we conducted an empirical study to understand silent failures through the lens of developers' practice of rerunning successful jobs in industrial contexts. 
\added{Our industrial setting at TELUS provides unique advantages for studying this phenomenon. Silent failures have been explicitly reported by developers across TELUS's heterogeneous software portfolio, which spans SDN controllers, backend APIs, and web applications across 13 programming languages. At TELUS, developers routinely rerun successful jobs when they suspect a silent failure has occurred.  Direct access to these practitioners and to 6 years of CI data across 81 projects enables us to study silent failures in context, unlike OSS-based studies that often lack developer insights. We further strengthen external validity through a replication on OSS projects and a qualitative analysis of public GitLab issues.} \added{In this study}, \added{we leverage the practice of rerunning successful jobs to uncover the influential factors, root causes, and practical solutions associated with silent failures in CI.}

\section{Threats to Validity}
\label{sec:threats}

\textbf{Construct Validity}. Threats to construct validity concern the degree to which our study measures what we claim to be analyzing. The most important threat of this kind is the identification of silent job failures, which we have studied in this paper, through successful jobs that have been rerun. Our main hypothesis is that developers rerun successful jobs only when these jobs have not fully or correctly completed their intended task. Such a hypothesis disregards cases where a truly successful job is still rerun, such as, for instance, the rerun of a deployment job after a scheduled environment cleanup. Nevertheless, TELUS engineers have confirmed our hypothesis in the studied industrial context, and we expect these cases to be very rare in other contexts since rerunning successful jobs does not align with effective CI practices \cite{elazhary_uncovering_2022}.

\added{To validate our proxy assumption that rerunning successful jobs signals silent failures, we manually analyzed a random sample of 100 rerun-initiating successful jobs from the TELUS dataset. For each job, we retrieved via the GitLab API the execution log, job metadata (i.e., name, stage, duration, artifacts, runner information, \texttt{\small allow\_failure} setting), commit details, pipeline context (sibling job statuses), and the same data for the corresponding rerun. Each job was initially classified using an LLM-as-a-judge approach~\cite{gu_survey_2025}, where the full job metadata, logs, and pipeline context were provided to Claude Sonnet 4.6 via a structured prompt included in our replication package~\cite{aidasso_replication_2025}. The first author then manually reviewed all 100 classifications, cross-referencing with feedback from TELUS engineers regarding known job behaviors and rerun practices. Jobs were classified into four categories: \emph{confirmed silent failure} (direct evidence such as error messages, missing artifacts, or ignored exit codes), \emph{likely silent failure} (circumstantial evidence such as missing logs due to documented runner crashes or trigger jobs reporting success before downstream pipeline completion), \emph{likely not a silent failure} (developer reran out of caution, i.e., defensively due to flaky external dependencies or unobservable configuration changes), or \emph{inconclusive}.}

\added{Of the 100 sampled jobs, 19 were classified as confirmed and 39 as likely silent failures, yielding a combined rate of 58\% silent failures. The remaining 36 jobs were classified as likely not a silent failure and 6 jobs as inconclusive. The distribution of themes across the 58 silent failures aligns with the RQ\textsubscript{3} taxonomy, with post-deployment issues (T7, 29 cases) and infrastructure configuration issues (T4, 10 cases) being the most prevalent. These results confirm that the majority of rerun-initiating successful jobs correspond to actual silent failures. We acknowledge that the proxy is not perfect and, accordingly, present our findings as factors associated with the rerun of successful jobs, of which the majority (58\%) are attributable to silent failures. The complete validation dataset and results are included in our replication package~\cite{aidasso_replication_2025}.}

Another threat to construct validity comes from our use of open coding, which involves manual inspection and subjective judgment. In RQ\textsubscript{2}, we derived job types from job names using regex-based classification. While this process may introduce errors, we carefully selected keywords that clearly indicate the job’s category, thereby reducing the risk of mislabeling. To mitigate this threat in RQ\textsubscript{3}, where the assignment of themes was more interpretive, the first two researchers independently classified the GitLab issues. We then calculated an agreement score of 0.913, which shows almost perfect agreement. We make all related artifacts publicly available \cite{aidasso_replication_2025}, enabling verification and supporting replication of our results.

\added{Finally, the 92 public issues (1,199 comments) analyzed in RQ\textsubscript{3} were retrieved exclusively from the GitLab issue tracker, which may over-represent problems specific to the GitLab ecosystem and issues that are sufficiently severe or visible to be publicly reported. Silent failures that are resolved internally without a public report are not captured. Nevertheless, publicly reported issues represent real problems encountered by practitioners, lending practical relevance to our findings. Furthermore, the platform-agnostic nature of most identified themes (e.g., ignored exit codes, caching errors, artifact issues) suggests broader applicability beyond GitLab.}

\textbf{Internal Validity}. The main internal threat to validity lies in the omission of confounding factors that may influence our interpretation. Indeed, we did not explore all the factors that could be associated with the re-execution of successful jobs, such as the CI server characteristics or variables indicating whether the job used cache. Such variables could better explain the likelihood of rerunning successful jobs. However, these data are either not openly available or require feature engineering too specific to the CI service, which would pose a threat to generalizability. \added{Besides, code change intent before reruns could influence the prevalence of silent failures and warrants investigation in future commit-level analyses.} Nevertheless, we selected 14 primary factors (resulting in 32 independent variables) across three dimensions based on existing literature studying factors associated with build failures \cite{islam_insights_2017, rausch_empirical_2017, dimitropoulos_continuous_2017, barrak_why_2021} and build performance \cite{weeraddana_characterizing_2024, ghaleb_empirical_2019, silva_what_2023, kola-olawuyi_impact_2024}. Also, the purpose of this study is to highlight the factors associated with rerunning successful jobs so as to raise awareness within industrial development teams. We encourage future research to investigate the causal relationships with these factors. 

On the other hand, the variable selection following correlation analysis may have influenced the results of our mixed-effects models. To mitigate this concern, we provided a detailed rationale for the choice of variable in each pair of correlated variables. Moreover, we provide a replication package \cite{aidasso_replication_2025} to facilitate the reproducibility of our results.

\textbf{External Validity}. We built our mixed-effects models with data from projects using GitLab CI. As such, \removed{our findings may not generalize to other CI services. However, the phenomenon of silent job failures is not inherently specific to any CI service, although we argue that it might be more prevalent in industrial settings where CI services and tools (e.g., artifact repository) are self-hosted.}
\added{some of our findings are inherently tied to GitLab CI's design, such as the specific artifact and cache handling mechanisms (e.g., \texttt{artifacts:when} semantics, cache key strategies) and infrastructure issues specific to certain runner configurations (e.g., Windows-based runners). However, the core findings of this study extend beyond any single CI platform. Silent failures, by nature, arise from the gap between a job's reported status and its actual outcome, a problem common to all CI systems, although we argue that silent failures might be more prevalent in industrial settings where CI services and tools (e.g., artifact repositories) are self-hosted. Also, testing and artifact-producing jobs exist on every CI platform and remain susceptible to silent issues due to their reliance on external resources such as caches, artifact stores, and test frameworks. Likewise, root causes such as ignored exit codes, scripting pitfalls, and misconfigured infrastructure are platform-agnostic.} Furthermore, \removed{the variables we defined are statically computable and CI service-agnostic}\added{the variables used in our study (Table~\ref{tab:features}) are CI service-agnostic and computable from metadata available in most CI platforms}. To facilitate replication on other CI services such as Jenkins or GitHub Actions, we have made our replication package publicly available~\cite{aidasso_replication_2025}.

In addition, the focus on industrial projects at TELUS poses another threat to external validity. To address this, we analyzed 81 projects that vary in size, age, and programming language (13 in total, including widely used ones such as Python, JavaScript, and Go). These projects serve diverse purposes, ranging from backend systems (e.g., SDN controllers, APIs) to front-end applications (e.g., business intelligence dashboards and web portals). TELUS also makes extensive use of widely adopted open-source tools such as Docker and SonarQube. We believe that such diversity of project context and tools enhances the generalizability of our results beyond the context of this study. Additional details on the studied projects, including their purpose, age, dominant language, number of commits, jobs, etc., are provided in our {Online Appendix B}.\cref{note:online_appendix} We have openly released the industrial dataset of 142,387 jobs to support open science. \added{We also note that excluding projects with fewer than 50 commits may bias the findings toward more mature projects with established CI practices.}

Nonetheless, we conducted a replication study on 198,146 jobs collected across 8 popular GitLab open-source projects to ensure external validity of our findings. This dataset has been extracted from the \textsc{GlBuilds} dataset\footnote{\url{https://doi.org/10.6084/m9.figshare.27641388.v1}} introduced previously in \cite{aidasso_towards_2025}. \added{Key replication results summarized in RQ\textsubscript{1} and RQ\textsubscript{2} (Sections~\ref{sec:rq1} and~\ref{sec:rq2}), confirmed many of our findings, including the distinction between silent failures and true successes (AUC ROC of 97\%, AUC PR of 85\%) and the influence of factors such as testing and static code analysis jobs.} Complete results and comparative discussions are provided in our Online Appendix~F\cref{note:online_appendix}.

\section{Conclusion}
\label{sec:conclusion}

One major challenge in CI systems is the unreliability of pipeline outcomes, where job failures or successes may not reflect the actual state of job execution. While prior work~\cite{aidasso_diagnosis_2025, aidasso_efficient_2025, olewicki_towards_2022, lampel_when_2021, durieux_empirical_2020} has focused on intermittent job failures, the problem of false job successes has been largely overlooked. These ``silent failures'' are often difficult to trace using logs and are potentially more prevalent in industrial contexts managing self-hosted CI infrastructure. In this paper, we investigate silent failures through the lens of successful jobs that were later rerun by developers at a large telecommunications organization. \added{A manual validation of 100 sampled reruns confirms that the majority (58\%) correspond to actual silent failures.} We found that over half of reruns follow successful jobs, and often occur after delays of up to several days, suggesting delayed detection of silent issues. \added{These reruns account for 17.3\% of the total CI compute budget, representing substantial waste for organizations.} Reruns of successful jobs are more likely for code or image build, testing and static code analysis jobs, and in projects dominated by scripting and infrastructure-as-code languages such as Shell and HCL. \added{A thematic analysis of 92 publicly reported issues identified 11 root cause categories, with artifact operation errors, caching errors, and ignored non-zero exit codes being the most prevalent. A cross-tabulation of these themes against job types confirms that build and testing jobs concentrate in the most frequent silent failure categories, aligning with their higher rerun odds in the quantitative analysis.} \added{Based on these findings, we provide concrete recommendations for practitioners, including post-execution verification steps, stricter exit code handling, and cache configuration reviews. We also identify CI configuration smells that produce fake successes and suggest that extending CI linting tools with detection rules for these patterns could prevent a subset of the silent failures we identified.}

\bibliographystyle{IEEEtran}
\bibliography{references}

\vspace{11pt}
\vfill\eject

\vspace{-33pt}
\begin{IEEEbiography}[{\includegraphics[width=1in,height=1.25in,clip,keepaspectratio]{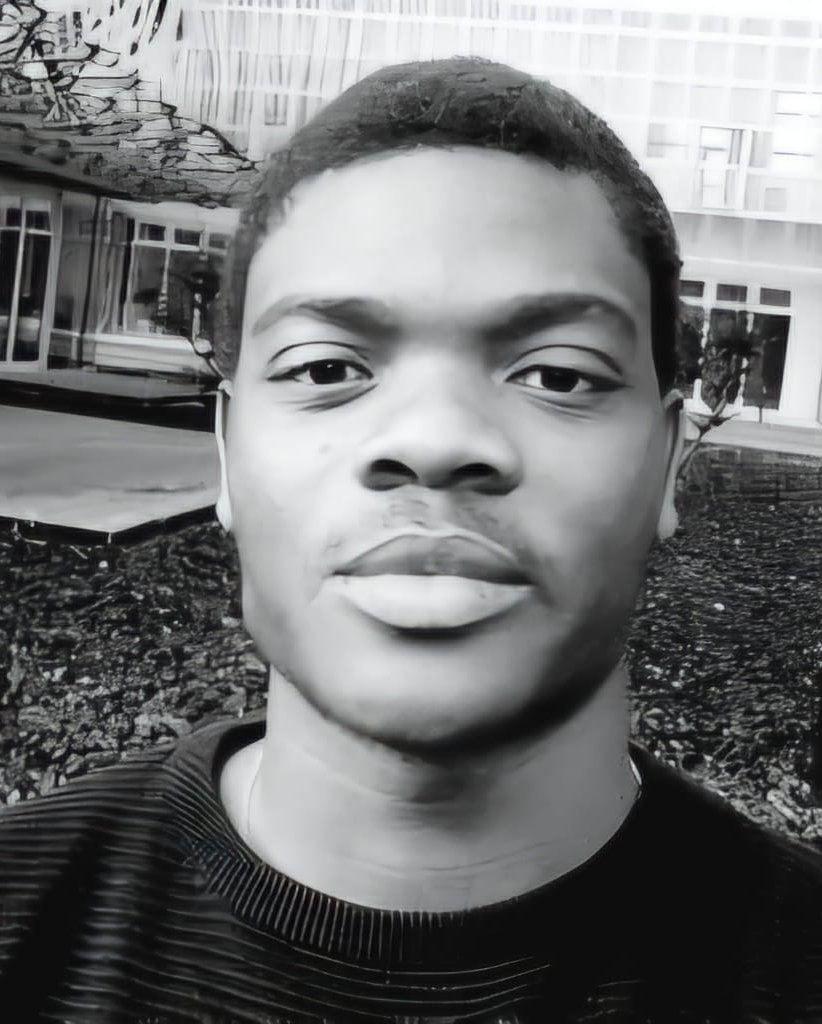}}]{Henri Aïdasso} is a Computer and Software Engineering PhD candidate at École de technologie supérieure (ÉTS, University of Quebec), Montreal, Canada, and a Mitacs Accelerate Fellow affiliated with TELUS. He received his M.Sc. in Big Data: Decision Intelligence and Machine Learning, graduating valedictorian from the University of Rennes 1 in France. He holds a B.Sc. in Computer Science from the University of Rennes 1 with the highest honours and also a B.Sc. in Computer Science applied to Management from the University of Abomey-Calavi, Benin, where he was awarded the Trophy of Excellence. His current research interests include applied machine learning, natural language processing, software evolution, software reliability, and intelligent DevOps.
\end{IEEEbiography}

\vspace{-25pt}
\begin{IEEEbiography}[{\includegraphics[width=1in,height=1.5in,clip,keepaspectratio]{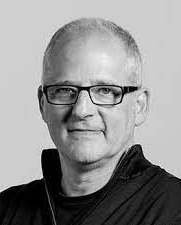}}]{Francis Bordeleau} is a Professor at École de technologie supérieure (ÉTS), Montreal, Canada, where he holds the ÉTS Kaloom-TELUS Industrial Research Chair in DevOps. He has over 25 years of experience in software and model-based engineering (MBE), including research, development, consulting, and collaboration with companies worldwide. Before joining ÉTS, he co-founded Cmind and was Product Manager at Ericsson (2013-2017), where his primary responsibilities included MBE and modeling tools. His research interests include model-based engineering, software processes, DevOps, and digital twins. He received his Ph.D. in Electrical Engineering from Carleton University, Ottawa, Canada.
\end{IEEEbiography}

\vspace{-25pt}
\begin{IEEEbiography}[{\includegraphics[width=1in,height=1.25in,clip,keepaspectratio]{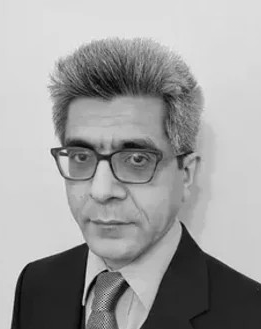}}]{Ali Tizghadam} is a Technology Fellow and Chief Automation Architect affiliated with the CTO office at TELUS. In his current position, he leads TELUS’s transition toward self-driving networks via network softwarization and cross-domain automation strategies. He is also a part-time Senior Researcher and Lecturer at the University of Toronto. He has over 25 years of experience in the telco industry, holding executive positions such as VP of Innovation, Director of R\&D, and startup founder. He obtained his Ph.D. in Electrical and Computer Engineering from the University of Toronto, Canada. His main research interests include autonomous networks, intent-driven automation, and intelligent DevOps.
\end{IEEEbiography}

\vfill

\end{document}